%% file: price_transit.tex
\newif\ifLNCS
\LNCStrue
\newif\ifnotLNCS
\notLNCSfalse

\ifLNCS
\RequirePackage{amsmath}
\fi
\documentclass[runningheads]{llncs}
\usepackage{graphicx}

\ifLNCS
\usepackage{makeidx}  
\fi

\usepackage{times}
\usepackage{xcolor}
\usepackage{soul}
\usepackage[utf8]{inputenc}

\input{macros}

\title{Transitions of Solutions and Their Efficiency}

\author{Gleb Polevoy\inst{1}} 
\institute{Paderborn University, Paderborn, Germany}

\begin{document}

\ifLNCS
%
%
\fi

\ifLNCS
\fi
\maketitle 



\input{abstract}

\input{introduction}
\input{model}
\input{eff_cc_transit}

\input{lim_transit}
\input{ext_smooth}

\input{eff_class_game}

\input{stable_trans_coord_game}
\input{alg}
\input{background}
\input{conclusion}




\bibliographystyle{splncs04}
\bibliography{library}

\newpage
\appendix
\input{omit_proofs}
\input{decomp_zero_sum_pot}

\end{document}

%% file: macros.tex
\usepackage[ruled]{algorithm2e}

\SetAlFnt{\small}
\SetAlCapFnt{\small}
\SetAlCapNameFnt{\small}
\SetAlCapHSkip{0pt}

\usepackage{url}

\usepackage{latexsym,amsmath,amssymb}
\ifnotLNCS
\usepackage{amsthm}
\fi
\usepackage{graphicx}

\usepackage{ifthen}
\usepackage{mdwlist,paralist}

\usepackage{comment}
\excludecomment{longversion}

\usepackage{epsfig}
\usepackage{epsf}

\usepackage[geometry]{ifsym}
\usepackage{rotating}

\usepackage{float}

\newboolean{Longversion} \setboolean{Longversion}{false}

\newcommand*{\eqns}[1]{Eq.~#1}
\newcommand*{\eqnsref}[1]{\eqns{\eqref{#1}}} 

\newcommand*{\chapt}[1]{Chapter~#1}
\newcommand*{\sectn}[1]{Section~#1}
\newcommand*{\sectnref}[1]{\sectn{\ref{#1}}} 
\newcommand*{\fig}[1]{Figure~#1}
\newcommand*{\figref}[1]{\fig{\ref{#1}}} 

\newcommand{\anote}[1]{}

\newcommand{\argmin}{\ensuremath{\mbox{argmin}}} 
 
\newcommand{\set}[1]{\left\{ #1 \right\}} 
 
\newcommand{\paren}[1]{\left( #1 \right)}

\newcommand{\abs}[1]{\left| #1 \right|} 
\newcommand{\ceil}[1]{\left\lceil {#1} \right\rceil} 
\newcommand{\floor}[1]{\left\lfloor {#1} \right\rfloor}

\newcommand{\reals}{\mathbb{R}}
\newcommand*{\realsP}{{\ensuremath{\mathbb{R}_{+}}}}

\newcommand*{\NP}{\textrm{NP}}

\newcommand{\calC}{\mathcal{C}}

\newcommand{\calP}{\mathcal{P}}

\newcommand*{\defas}{\ensuremath{\stackrel{\rm \Delta}{=}}}

\DeclareMathOperator*{\Neighb}{N}

\ifnotLNCS
\newtheorem{theorem}{Theorem}
\fi
\newtheorem{defin}{Definition}
\ifnotLNCS
\newtheorem{remark}{Remark}
\newtheorem{lemma}{Lemma}
\newtheorem{proposition}{Proposition}

\newtheorem{corollary}{Corollary}
\fi
\newtheorem{observation}{Observation}
\ifnotLNCS
\newtheorem{example}{Example}

\newenvironment{definition}{\defin \em }
\fi

\newcommand{\defined}[1]{{\bf\emph{#1}}}

\newcommand{\NE}{\text{NE}\xspace} 

\DeclareMathOperator{\poa}{PoA} 
\DeclareMathOperator{\pota}{PoTA} 
\DeclareMathOperator{\poccta}{Po\cct{}A} 
\DeclareMathOperator{\posta}{Po\stablT{}A} 
\DeclareMathOperator{\pos}{PoS} 
\DeclareMathOperator{\pots}{PoTS} 
\DeclareMathOperator{\poccts}{Po\cct{}S} 
\DeclareMathOperator{\posts}{Po\stablT{}S} 

\DeclareMathOperator{\sw}{SW} 
\DeclareMathOperator{\scost}{SC} 
\DeclareMathOperator{\BR}{BR} 

\DeclareMathOperator{\cct}{CT} 
\newcommand{\stablT}{ST} 



%% file: abstract.tex
\begin{abstract}
We broaden the basis of non-cooperative game theory by
considering miscoordination on a solution concept.
For any solution concept,
we extend the solution set of a strategic-form game to a transition set.
This set contains profiles where various agents simultaneously follow different
solutions, e.g.~different Nash equilibria. This models
the fact that in practice, complicated agents
are rarely perfectly coordinated on the same equilibrium.
We define two efficiency measures, called the price of 
transition anarchy and stability, 
and bound them.
We also refine the notion of transition
to the notion of limited transition, where only a limited number of solutions
is simultaneously played, and to stable transitions, which allow for only
minor lack of coordination.
We compare the above mentioned efficiency measures and
bound the efficiency of transitions in important cases, including the important cases
of constant-sum and potential games, which span the set of finite games
with the same number of strategies for each agent.
We also prove tight efficiency bounds for routing games and
coordination games on graphs. 
Finally, we study algorithms to find the transition degree required
to make a given profile a transition, or to render all the profiles
transitions.
We conclude 
that for the sake of efficiency, it is crucial to avoid uncoordinated
transitions, besides certain cases, such as constant-sum games, identical utility games,
some types of routing games, limited transitions in potential games,
and stable transitions in coordination games.
\end{abstract}

%% file: introduction.tex
\section{Introduction}

Many solution concepts, such as Nash equilibria,
impose unrealistic assumptions.
First,
much of non-cooperative game theory studies existence and properties of solutions
that describe either steady states
or rationally deduced profiles%
~\cite[\sectn{1.5}]{OsborneRubinstein94}. However, 
steady states require remembering history,
while rational deduction of, say, Nash equilibria (\NE) often requires belief assumptions that are not
easy to satisfy, as shown in~\cite[\sectn{5.4}]{OsborneRubinstein94}
and~\cite[\chapt{4}]{Perea2012}. This problem 
suggests considering profiles requiring less belief assumptions.

An important practical problem with assuming a single solution (instance of the given solution concept) is that people
often move from one modus operandi (equilibrium) to another not 
simultaneously. For instance, 
some people may become economically and politically active, while the others
are still used to the old life style with less personal power and responsibility.
Also the changes in the concept of marriage make
various people have different perceptions of marriage.
Various road networks allow for different
equilibrium flows~\cite{Braess1968}, and different 
drivers choose their path according to their own expectations. 
Assuming a single solution is also problematic when the agents are
not coordinated in their actions,
as described in~\cite[\chapt{12}]{ManktelowChung2004}.
These examples suggest how to modify the definition of a solution
to describe the actual behavior.

The examples of lack of synchronization on a solution
call for formalizing the scenario
of various agents taking actions from not necessarily the same solution
and analysing the efficiency repercussions of such actions.
Since we are not aware of any analysis of profiles
that can be seen as transitions from one solution to another,
we define the natural meta-solution concept of transition, 
which relaxes the belief and coordination assumptions.
We also refine this broad
meta-solution concept, to make it realistic, and study the social
welfare when the agents play accordingly.
We model a static situation of different players playing various
solutions, without analysing the dynamics that may accompany playing
the transitions. Such dynamics would depend on what the agents know and
how they decide on their actions and are beyond our scope.
Given (any) solution set, a \defined{transition}
is a profile where every agent acts according to one of the solutions (e.g., one of the Nash equilibria),
but not necessary everyone follows the same solution.
This also remedies the above-mentioned problem of belief assumptions required
for solution concepts such as Nash equilibrium, since here, a player
does not need to know what the others do or used to do.

A transition profile exists by
definition, provided at least one solution exists, which we
always assume. We study
the social efficiency of the worst (best) transition profiles, relatively
to the maximum
possible efficiency. We call this ratio
\defined{the price of transition anarchy (stability)},
inspired by the celebrated price of
anarchy~\cite{KP99,KoutsoupiasPapadimitriou2009} and
stability~\cite{SchulzStierMoses2003,AnshelevichDasGuptaKleinbergTardosWexlerRoughgarden04}.
Since the concept of a transition is more
realistic than its underlying solution concept, so is
the price of transition anarchy, and it models the worst decrease
caused by lack of knowledge or by miscoordination. The price of transition stability
can be seen as the socially best situation where all the agents
think they are playing an equilibrium. This can result from a mischievous
coordination, when everyone thinks he is proposed an equilibrium strategy,
but the whole profile is not an equilibrium. The best transition is
also the best profile that can result from lack of coordination.

Sometimes, agents can only choose between few solutions.
For example, only a few equilibrium flows are well known to agents
or suggested by the routing systems.
In order to model partial incoordination,
we refine the notion of transition and our efficiency results
to the so called \defined{limited transitions} that describe
agents following a predefined number of solutions, and
define the appropriate notions of their efficiency.
We also model that the most realistic transitions require every
sub-optimally acting player be rendered optimal by some other
player's improvement, since otherwise, this sub-optimally acting
player would have improved. This also represents that lack of
clarity as to what the other players might do. We call such profiles
\defined{stable transitions} and analyse their structure and
improved efficiency for coordination games.

This paper considers only pure solutions.
We define the notion of transition
and the corresponding efficiency measures in \sectnref{Sec:model}.
We refine the notion of transition by defining limited transitions in \sectnref{Sec:model:lim_transit}
and stable transitions in \sectnref{Sec:model:stabl_transit}.
We provide general
efficiency bounds in \sectnref{Sec:eff_cc_transit}.
Additional bounds for the central solution concept of Nash equilibrium
appear in \sectnref{Sec:ext_smooth}, closely analysing the important cases of
constant-sum and potential games. Since a finite game with the same number
of strategies per each agent can be decomposed to zero-sum and
a potential game, this allows to prove a bound on a general game as well,
a bound that depends on how close the given game is to a potential game.
We prove tight efficiency guarantees for routing games 
in \sectnref{Sec:rout_game} and study stable transitions in \sectnref{Sec:stable_trans_coord_game}.
We then study algorithmic questions in \sectnref{Sec:alg_comp}.
The related literature is described in \sectnref{Sec:background} and
we conclude in \sectnref{Sec:conclusion}. Appendix
\sectnref{Sec:omit_proofs_etc} presents the proofs and examples omitted
from the body of the paper, and \sectnref{Sec:decomp_zero_sum_pot} describes how to
use the decomposition of a game into zero-sum and potential games to
bound its efficiency. 

This work broadens the basis of non-cooperative game theory by defining practically important extensions of solutions to games
and bounding their efficiency, thereby laying the basis for realistic analysis of
important interactions.

%% file: model.tex
\section{Model}\label{Sec:model}

Consider a strategic form \defined{game}
$G = (N, S = S_1 \times S_2 \times \ldots \times S_n, (u_i)_{i = 1, \ldots, n})$,
where $N = \set{1, \ldots, n}$ is the set of
agents, $S_i$ is agent $i$'s strategy set and
$u_i \colon S \to \reals$ is agent~$i$'s utility function.
The \defined{solutions}, forming a \defined{solution set}, are a set of strategy profiles $D \subseteq S$,
defined for a game by a rule called \defined{a solution concept}.
For example, one of the most famous solution sets is the \defined{the Nash equilibria},
denoted $\NE$,\footnote{In text, \NE{} also abbreviates the words
``Nash  equilibrium(a)'', when no confusion with the notation of the set arises.}
which constitute the following set of
strategy profiles. A Nash equilibrium~\cite{Nash51} of $G$ is a strategy
profile $s = (s_1, \ldots, s_n) \in S_1 \times \ldots \times S_n = S$
such that
\begin{equation}
\forall i \in N, \forall s_i' \in S_i: u_i(s) \geq u_i(s_i', s_{-i}),
\end{equation}
where $s_{-i} \defas (s_1, \ldots, s_{i - 1}, s_{i + 1}, s_n)$.

We now relax a solution concept, expressing the idea
that some agents act according to one solution, while others act
according to, perhaps, different solutions. This models movement between
solutions or incoordination on a solution.
\begin{defin}\label{def:transit}
Given a solution set $D \subseteq S$, define a
\defined{transition} as any profile $s = (s_1, \ldots, s_n) \in S$ such
that for each $i \in N$, there exists a solution
$d(s, i) = (d_1, \ldots, d_n) \in D$, such that $s_i = d_i$.
Denote the set of all the transitions to be $T(D) \subseteq S$, the
\defined{transition set}.
\end{defin}
A set of profiles is a transition set (of some solution set) if and only if it is a Cartesian
product. Indeed,
by definition, $T(D)$ is the Cartesian product of the projections
of the solution set,
$D_i = \set{d_i | \exists d = (d_1, \ldots, d_n) \in D}$.
For example, in a symmetric game, all the projections of the Nash equilibria set
are identical.
In the other direction, any Cartesian product is a transition set of
itself.

By definition,
$D \subseteq T(D)$, since a player playing a transition
can pick an action from any solution, regardless the other players.
Taking transitions is idempotent, i.e.~if we define a new solution set as $T(D)$,
then $T(T(D)) = T(D)$.

$T(D) = D$ if and only if
$D$ is a product set. For example, for the game of
Matrix \eqref{eq:coord_game_ineff}, 
$\NE = \set{(I, 1),(II, 2)}$, while $T(\NE) = S$.
On the other hand, solution sets such as rational
choices~\cite[Definition~{2.4.5}]{Perea2012} are by definition products of
sets, and, therefore,
$T(\text{rational choices}) = \text{rational choices}$.

\begin{equation}
\begin{array}{l|c|c|}
		& 1:	& 2:	\\
\hline
I:	&	(a, a)	&  (0, 0)	\\
\hline
II:	& (0, 0)	&  (a, a)	\\
\hline
\end{array}
\label{eq:coord_game_ineff}
\end{equation}%

$T(D) \neq \emptyset \iff 
D \neq \emptyset$, which we always assume.
Computing $T(D)$ means finding all the projections of
$D$. 

In order to measure efficiency,
call the total utility in a strategy profile $s \in S$ the
\defined{social welfare}, i.e.~$\sw(s) \defas \sum_{i \in N}{u_i(s)}$.
Inspired by the notions of the price of anarchy%
~\cite{KP99,KoutsoupiasPapadimitriou2009,Papadimitriou2001} and
the price of stability~\cite{SchulzStierMoses2003,AnshelevichDasGuptaKleinbergTardosWexlerRoughgarden04},
which compare the least or the largest possible social welfare in a
solution to the maximum possible social welfare, we compare the
least or the largest possible social welfare in a transition 
with the maximum possible social welfare.
Formally,\footnote{We reverse the usual ratio to work with prices between $0$ and $1$.}
$\poa \defas \frac{\min_{s \in D}{\sw(s)}}{\max_{s \in S}{\sw(s)}}$
and
$\pos \defas \frac{\max_{s \in D}{\sw(s)}}{\max_{s \in S}{\sw(s)}}$.
(For cost-minimization,
$\poa \defas \frac{\max_{s \in D}{\scost(s)}}{\min_{s \in S}{\scost(s)}}$
and
$\pos \defas \frac{\min_{s \in D}{\scost(s)}}{\min_{s \in S}{\scost(s)}}$,
where the social cost is defined as $\scost(s) \defas \sum_{i \in N}{c_i(s)}$.)
We now define
\begin{defin}\label{def:efficiency_measures}
Define \defined{the price of transition anarchy ($\pota$)} to be the
ratio of the least possible social welfare in a transition to the maximum
possible social welfare, i.e.~%
$\pota \defas \frac{\min_{s \in T(D)}{\sw(s)}}{\max_{s \in S}{\sw(s)}}$.
(In the case of cost-minimization,
$\pota \defas \frac{\max_{s \in T(D)}{\scost(s)}}{\min_{s \in S}{\scost(s)}}$.)

Let \defined{the price of transition stability ($\pots$)} be the
ratio of the largest possible social welfare in a transition to the maximum
possible social welfare, i.e.~%
$\pots \defas \frac{\max_{s \in T(D)}{\sw(s)}}{\max_{s \in S}{\sw(s)}}$.
(For cost-minimization, we define
$\pota \defas \frac{\min_{s \in T(D)}{\scost(s)}}{\min_{s \in S}{\scost(s)}}$.)

\end{defin}

If the minima or maxima are not always attained, they should be
replaced with infima or suprema, respectively. All the statements we prove
keep holding in this general case. This holds also for the refinements
in the following subsections, which model
the practical limitations on the complete
lack of coordination that transitions allow.

\subsection{Limited Transitions}\label{Sec:model:lim_transit}

The assumption that any player will merely act according to her strategy
in some solution regarding the other players is broad and therefore safe, but 
in
some situations, there are only few reasonable solutions to choose from,
allowing us refine the model and increase the predictive capability. 
%
Intuitively, transitions allow for a complete lack of coordination
in choosing the solution according to which to act, while choosing a
complete solution allows for no lack of coordination here, since
everyone is coordinated on the chosen solution. We now refine
these too concepts to allow for a partial lack of coordination.
\begin{defin}\label{def:transit:lim}
Let $m$ be a natural number.
Given a solution set $D \subseteq S$, define an
\defined{$m$-transition} as any profile $s = (s_1, \ldots, s_n) \in S$ such
that there exist at most $m$ solutions $D(s) \subseteq D$, perhaps
dependent on $s$, and
for each $i \in N$, there exists a solution
$d(s, i) = (d_1, \ldots, d_n) \in D(s)$, 
such that $s_i = d_i$.
Denote the set of all the $m$-transitions to be $T(D, m) \subseteq T(D) \subseteq S$, the
\defined{$m$-transition set}.
The minimum $m$ such that a given transition is an $m$-transition is
called the \defined{transition degree} of a transition.
\end{defin}
We have $D \subseteq T(D, m) \subseteq T(D)$, 
and fully written, 
$D = T(D, 1) \subseteq T(D, 2) \subseteq \ldots \subseteq T(D, n - 1) \subseteq T(D, n) = T(D)$.
Unlike with the transition set $T(D)$, the $m$-transition $T(D, m)$ is not
idempotent, namely $T(T(D, m), m') = T(D, m \cdot m')$. 

\anote{Another refinement of a transition can be relative to a player. We
can bound the number of players that play not according to any of the
solutions that are played by the player. This is relevant to the case
of several independent games.}

This definition automatically allows to refine the price of transition
anarchy and stability to the price of $m$-transition anarchy and
stability. Namely, $m-\pota \defas \frac{\min_{s \in T(D, m)}{\sw(s)}}{\max_{s \in S}{\sw(s)}}$
and $m-\pots \defas \frac{\max_{s \in T(D, m)}{\sw(s)}}{\max_{s \in S}{\sw(s)}}$.
(For cost-minimization,
$m-\pota \defas \frac{\max_{s \in T(D, m)}{\scost(s)}}{\min_{s \in S}{\scost(s)}}$
and $m-\pots \defas \frac{\min_{s \in T(D, m)}{\scost(s)}}{\min_{s \in S}{\scost(s)}}$.)

\subsection{Stable Transitions}\label{Sec:model:stabl_transit}

Despite the applicability of the following definitions to
any solution concepts, we consider stable transitions of Nash
equilibria. 
A transition allows for any combination of solutions.
Consider the expressive example of coordinating choices
of the (graph) coordination game~\cite{AptdeKeijzerRahnSchaeferSimon2017}.
The players are identifies with the nodes, and the strategies of
a node are the colours from its own predefined set of colours.
The utility of a node is the number of its neighbours that
have chosen the same colour as the node itself has. If
every node has only two possible colours,
$\set{r, b}$, then regardless the topology of the underlying graph,
there exist at least the following Nash equilibria: everyone
picks $r$, or everyone picks $b$. Thus, even the set of $2$-transitions
contains all the possible strategy profiles, which is very permissive.

In order to improve the predictive power of the solution set in this and similar cases,
let us restrict the transitions to be realistic. Ex-ante, a
player may prefer only the transitions where she plays an equilibrium that yields her most
utility, while ex-post, a transition where a player can increase
her utility, regardless the others' unilateral improvements,
seems unstable. The first restriction is less practical, since you can
play an equilibrium where you're not well off, provided you believe
that is going to be played. Additionally, in the example of the bi-colour
coordination game above, both mentioned Nash equilibria yield equal
maximum utility to everyone, so this refinement would not restrict the
transition set there anyway. Instead, we now model the ex-post limitation.

\begin{defin}\label{def:transit:stabl}
Given a solution set $D \subseteq S$, a \defined{stable transition}
$s$ is a transition of $D$ where for any not best-responding player~$i$
there is a non best-responding player~$j$ that has a best response that
would render $i$'s original strategy to be a best response. In formulas,
\begin{eqnarray*}
\forall i \in N, \text{such that } s_i \notin \BR(s_{-i}),\\
\exists j \in N, \text{such that } s_j \notin \BR(s_{-j})
\text{ and } \exists \hat{s_j} \in \BR(s_{-j}), \text{such that }
s_i \in \BR(\hat{s_j}, s_{-\set{i, j}}).
\end{eqnarray*}

We call the set of all the stable transitions the
\defined{stable transitions set},~$ST(D)$.
\end{defin}
This definition can be generalised to multiple such players~$j$, and
the results of the whole paper can by generalised accordingly. However,
this generalisation is no strengthening, namely 
$(m + 1)-SE(D) \not \subseteq m-ST(D)$.

We have $D \subset ST(D) \subset T(D) \subset S$. Like transitions,
applying stable transitions are idempotent, i.e.~$ST(ST(D)) = ST(D)$,
since $ST(ST(D)) \subseteq ST(D)$, and since no new constraints are
applied, we obtain the equality.

We immediately refine the efficiency concepts for the price of
stable transition anarchy~$\posta$ and stability~$\posts$, as follows.
$\posta \defas \frac{\min_{s \in ST(D)}{\sw(s)}}{\max_{s \in S}{\sw(s)}}$
and $\posts \defas \frac{\max_{s \in ST(D)}{\sw(s)}}{\max_{s \in S}{\sw(s)}}$.

%% file: eff_cc_transit.tex
\section{Bounding Efficiency of Transitions}\label{Sec:eff_cc_transit}

We start with an efficiency bound, requiring the transition to be 
both limited and stable. We assume in the next theorem that the players 
interact in pairs, so we first define a polymatrix game~\cite{Janovskaja1968}.
\begin{definition}
A \defined{polymatrix game} on players $N$ with strategy
sets $(S_i)_{i \in N}$ has the following utilities.
Given the matrices
$U_{i, j} \in \reals^{S_i \times S_j}$ for each $i \neq j \in N$, we have
$u_i \defas \sum_{j \in N \setminus\set{i}} {U_{i, j}(s_i, s_j) }$.
\end{definition}
We also assume a special symmetry of the polymatrix game, which
is not the standard one, but is rather defined as follows.
\begin{definition}
A polymatrix game is \defined{symmetric} if 
\begin{enumerate}\label{def:sym}
	\item \label{def:sym:get_eq}
	for every player~$i \in N$,
	$U_{i, j} = U_{i, k}, \forall j, k \in N \setminus \set{i}$;
	\item	\label{def:sym:mon_sw_ut}
	and for all players~$i \in N$, the following monotonicity holds:
	$\sw(s) \geq \sw(t) \Rightarrow u_i(s) \geq u_i(t)$.
\end{enumerate}
\end{definition}
In words, first we fix a player, and then all the other players contribute
to her equally. 
Second, we demand that if one profile is socially better than another 
profile, than so relate also the utilities of each player in those profiles.
We also define a strategy profile $s \in S$ \defined{symmetric}, 
if $s_1 = \ldots = s_n$, and a set of profiles $D \subset S$ is 
\defined{symmetric} if every $s \in D$ is symmetric.

We will need the natural regularity condition on a polymatrix game, ensuring 
that in a transition, fixing some constituent solution~$s$, no player's 
unilateral deviation counted twice increases another player's utility by more 
than what other player received in total from all the players not playing $s$.
\begin{definition}
A polymatrix game 
$G = (N, S = S_1 \times S_2 \times \ldots \times S_n, U_{i, j} \in \reals^{S_i \times S_j}$
is called \defined{regular}, if the following holds for any transition $t$.
Denote the players playing some solution $d$ in $t$, by $P(d)$, namely
$P(d) \defas \set{i \in N : t_i = d_i}$. 
Then, for any solution $s$, any player~$j \in N$
and any player $i \neq P(s)$ there holds
$\sum_{k \neq (P(s) \cup \set{i})}{U_{i, k}(s_i, t_{k})}
\geq 2\max_{x_i \in S_i, y_j \in S_j}\set{U_{i, j}(x_i, y_j, t_{-\set{i, j}})}$.
\end{definition}
We are now ready to formulate and prove the theorem.
\begin{theorem}\label{The:sym_polym_limit_stabl_bound}
Given any symmetric solution set $D \subset S$ of
a nonnegative polymatrix symmetric regular game, 
$m-\posta \geq \poa / m$.
\end{theorem}

We first describe the main ideas of the proof:
\begin{description}
\item[Single users to $\sw$:] It is enough to prove that any player's utility
in a transition is at least $1 / m$ from her utility in some constituent
solution, since part~\ref{def:sym:mon_sw_ut} of stability allows carrying 
this over to social welfare.

\item[$m$-limited and symmetry imply the claim, if the player best responds:] 
The fact that we consider an $m$-limited transition and assume symmetry implies 
that any player's utility in that 
transition is at least $1 / m$ of her utility in some constituent
solution, if she best responds.

\item[Otherwise, stability and regularity help:]
If the player in question does not best respond, than the stability of the
transition implies that player best responds to a slightly perturbed profile,
and regularity allows us conclude the claim.
\end{description}

\begin{proof}
Fix any transition $t$ and any player~$i \in N$. It suffices to prove 
that $u_i(t) \geq (1 / m) u_i(s(i))$ for some solution $s(i) \in D$,
perhaps depending on~$i$. 
Indeed, then,
$\sw(t) = \sum_{\in N}{u_i(t)} \geq (1 / m) \sum_{i \in N}{u_i(s(i))}$,
and part~\ref{def:sym:mon_sw_ut} of symmetry will imply that 
$(1 / m) \sum_{i \in N}{u_i(s(i))} \geq (1 / m) \sum_{i \in N}{u_i(\hat{s})}$,
where $\hat{s} \defas \argmin_{i \in N}{\sw(s(i))}$. Finally,
recall that $(1 / m) \sum_{i \in N}{u_i(\hat{s})} = (1 / m) \sw(\hat{s})$.

Since $t \in T(D, m)$, namely an $m$-transition, $D$ is symmetric and $G$ is
a nonnegative polymatrix symmetric game (here part~\ref{def:sym:get_eq} 
of symmetry matters), there is a solution $s$, such that 
players~$P(s)$ contribute to $i$ at least $1 / m$ of $u_i(s)$, 
if $i$ plays $s_i$. If $u_i(t) \geq (1 / m) u_i(s)$ holds as well, 
we are done.

Now assume to the contrary that $u_i(t) < (1 / m) u_i(s)$. Then,
$t_i \not \in \BR(t_{-i})$, so we may employ the stability of 
the transition. Namely, $t \in ST(D) \Rightarrow$
$\exists j \in N, \text{such that } t_j \notin \BR(t_{-j})
\text{ and } \exists \hat{t_j} \in \BR(t_{-j}), \text{such that }
t_i \in \BR(\hat{t_j}, t_{-\set{i, j}})$.
We may assume $i \not \in P(s)$, since otherwise we
would have $u_i(t) \geq (1 / m) u_i(s)$, as
the game is nonnegative and polymatrix.
Now, the best $u_i$ player $i$ can achieve against $t_{-i}$ is at least
$\sum_{k \neq (P(s) \cup \set{i})}{U_{i, k}(s_i, t_{k})} + (1 / m) u_i(s)$.
Therefore, the highest attainable $u_i$ against $(t_{-\set{i, j}}, \hat{t_j})$
is at least 
$\sum_{k \neq (P(s) \cup \set{i})}{U_{i, k}(s_i, t_{k})} + (1 / m) u_i(s) 
- \max_{x_i \in S_i, y_j \in S_j}\set{U_{i, j}(x_i, y_j, t_{-\set{i, j}})}$,
to account for the disturbance of $j$'s deviation from $t_j$ to $\hat{t_j}$.
Since $t_i \in \BR(\hat{t_j}, t_{-\set{i, j}})$, we conclude that
$u_i(t_{-j}, \hat{t_j}) \geq \sum_{k \neq (P(s) \cup \set{i})}{U_{i, k}(s_i, t_{k})} + (1 / m) u_i(s) 
- \max_{x_i \in S_i, y_j \in S_j}\set{U_{i, j}(x_i, y_j, t_{-\set{i, j}})}$.
$\Rightarrow u_i(t) \geq \sum_{k \neq (P(s) \cup \set{i})}{U_{i, k}(s_i, t_{k})} + (1 / m) u_i(s) 
- 2\max_{x_i \in S_i, y_j \in S_j}\set{U_{i, j}(x_i, y_j, t_{-\set{i, j}})}$,
accounting once more for the disturbance of $j$'s deviation from $t_j$ to $\hat{t_j}$.
Finally, regularity implies the last expression is at least $(1 / m) u_i(s)$,
contradictory to the assumption $u_i(t) < (1 / m) u_i(s)$.
\end{proof}

Since we define transitions as combinations of solutions, it is natural
to ask whether there is a connection between the efficiency of
solution and that of transitions, namely, between the prices of anarchy
and stability and 
the prices
of transition anarchy and stability. 
Since the price of anarchy considers a worst possible solution, and the
price of stability considers the best possible one, we conclude:
\begin{observation}\label{obs:po_bound_nat}
\begin{eqnarray*}
\pota \leq \poa, \quad 
\pots \geq \pos\
\end{eqnarray*}
\end{observation}
The simple proof appears in the appendix.

In the opposite directions,
since transitions allow for arbitrary
picking of solutions by various players, there seems to be no reason for
lower (upper) guarantees on the price of transition anarchy (stability) to carry over
from the guarantees for the solutions, unless the game is insensitive to lack of
coordination. We first formally exemplify that guarantees indeed do not carry
over (Examples~\ref{ex:pota_can_dif_poa} and~\ref{ex:pots_can_dif_pos} in the appendix), and then define conditions that do make the guarantees carry over.

In the requirements
that would allow stating something about the efficiency of transitions
based on the efficiency of the underlying solutions, 
we can either
look at how the coordination of strategy choices influences \emph{social welfare},
or concentrate on the influence on \emph{individual utilities}.
We begin with the sensitivity of social welfare to
coordination between the strategic choices of the agents.
\begin{observation}\label{obs:po_bound}
Let $\alpha$ be at least $1$. For any game
$G = (N, S, (u_i)_{i = 1, \ldots, n})$
with a solution set $D \subseteq S$,
if $\min_{s \in T(D)}{\sw(s)} \geq \min_{t \in D}{\sw(t)} / \alpha$,
then
\begin{eqnarray}
\pota \geq \poa / \alpha.\label{eq:po_bound:a_low}
\end{eqnarray}

If $\max_{s \in T(D)}{\sw(s)} \leq \alpha \cdot \max_{t \in D}{\sw(t)}$,
then 
\begin{eqnarray}
\pots \leq \alpha \cdot \pos.\label{eq:po_bound:s:up}
\end{eqnarray}

This holds with equality
if and only if the condition inequality holds with equality.
\end{observation}
The immediate proof and a usage example appear in the appendix.

Observation~\ref{obs:po_bound} connects the efficiency of
a solution set and its transition set using the
dependency of the social welfare on coordination.
We now also present 
a direct condition on the influence of coordination of the strategies on an
individual player's utility that will allow connecting
the efficiencies of a solution set and its transition set.
\begin{defin}\label{def:alph_depend_util}
Player $i$'s utility over profile set $A \subseteq S$ is \defined{$\alpha$-lower (-upper) dependent on
coordination} if players taking strategies from various solutions can only
decrease (increase) $i$'s utility within the factor of $\alpha \geq 1$. Formally,
the lower dependency means
\begin{eqnarray}
\min_{s \in T(A)}{u_i(s)} \geq \min_{t \in A} u_i(t) / \alpha,
\label{eq:alph_low_depend_util}
\end{eqnarray}
and the upper dependency means
\begin{eqnarray}
\max_{s \in T(A)}{u_i(s)} \leq \alpha \cdot \max_{t \in A} u_i(t).
\label{eq:alph_up_depend_util}
\end{eqnarray}

\end{defin}

Having Definition~\ref{def:alph_depend_util} for all $i \in N$ over the solution set does not imply 
an $\alpha$ ratio between the transition efficiency measures and the
$\poa$ and $\pos$, because a
transition profile can be composed such that each agent's strategy is
taken from a solution where that agent obtains the lowest possible
utility, while that can be not the case for any solution profile. In order to be able to infer something about the $\pota$
and $\pots$, we need the following condition connecting
variations in an agent's utilities with variations in social welfare in various solutions.
\begin{defin}\label{def:bet_varied_util}
The utility of agent $i$ is \defined{$\beta$ varied over $A \subseteq S$} if
for all profiles $s, t$ in $A$,
\begin{eqnarray}
\sw(s) \geq \sw(t) \Rightarrow u_i(s) \geq u_i(t) / \beta.
\end{eqnarray}

\end{defin}

With Definitions~\ref{def:alph_depend_util} and~\ref{def:bet_varied_util} at hand,
we are finally able to connect the prices of transition anarchy and
stability with the prices of anarchy and stability.
\begin{proposition}\label{prop:po_bound_agentwise_cond}
Consider a game
$G = (N, S, (u_i)_{i = 1, \ldots, n})$
with a solution set $D \subseteq S$, such that over $D$,
the utility of every player~$i$ is $\beta$ varied
and $\alpha$-lower dependent on coordination, then
\begin{eqnarray}
\pota \geq \poa / (\alpha \beta).
\end{eqnarray}

If for every player~$i$, its utility over $D$
is $\beta$ varied
and $\alpha$-upper dependent on coordination, then
\begin{eqnarray}
\pots \leq \alpha \beta \pos.
\label{eq:po_bound_agentwise_cond:s_up}
\end{eqnarray}
\end{proposition}
The proof and usage examples appear in the appendix.
We have presented two kinds of conditions that allow connecting
the efficiency of transitions with the efficiency of their respective
solutions: a condition on the social welfare (Observation~\ref{obs:po_bound}) and a
condition on an agent's utility (Definition~\ref{def:alph_depend_util}).
%


%% file: lim_transit.tex
\subsection{Efficiency of Limited Transitions}\label{Sec:lim_transit}

\anote{We can generalize the coordination game example where the efficiency of the
transitions will change with $m$.}

Observations~\ref{obs:po_bound_nat} and~\ref{obs:po_bound} keep holding when
we say $m-\pota$ and $m-\pots$ instead of $\pota$ and $\pots$, respectively.
We can also generalize 
Definition~\ref{def:alph_depend_util} to the limited transition sets,
where the players may pick every time between at most $m$ different
solutions.
This generalization allows to correspondingly generalize
Proposition~\ref{prop:po_bound_agentwise_cond},
too. 

We now observe an implication of the influence of having more solutions
compose a transition on 
the social welfare.
\begin{observation}\label{obs:po_bound_limit}
In game~$G$ with a solution set $D \subseteq S$, if
$\min_{s \in T(A, i + 1)}{\sw(s)} \geq \min_{t \in T(A, i)}{\sw(t)} / \alpha_i$
holds for each $i = 1, \ldots, m - 1$, then
\begin{equation}
m-\pota \geq \poa / (\prod_{i = 1}^{m - 1}{\alpha_i}).
\end{equation}
In particular, $\pota \geq \poa / (\prod_{i = 1}^{n - 1}{\alpha_i})$.

If
$\max_{s \in T(A, i + 1)}{\sw(s)} \leq \alpha_i \max_{t \in T(A, i)}{\sw(t)}$
for each $i = 1, \ldots, m - 1$, then
\begin{equation}
m-\pots \leq (\prod_{i = 1}^{m - 1}{\alpha_i}) \pos,
\end{equation}
and, as a special case, $\pots \leq (\prod_{i = 1}^{n - 1}{\alpha_i}) \pos$.

Any of these statements holds as equality if and only if
the condition holds as equality
at each transition degree $1, \ldots, m - 1$.
\end{observation}
\begin{proof}
The proof 
follows by an inductive
application of the condition for $i = m - 1, \ldots, 1$, 
reducing the transition degree
in each step.
\end{proof}

Example~\ref{ex:match_strat} in the appendix demonstrates using this observation.
\newcounter{ex_match_strat}

The following definition brings a condition on how players adopting their
strategy from another solution can influence the individual utility of
a player.
\begin{defin}
Player $j$'s utility over $A$ at transition degree $m$ is
\defined{$\alpha$-lower-dependent on the transition degree} if
\begin{eqnarray}
\min_{s \in T(A, m + 1)}{u_j(s)} \geq \min_{t \in T(A, m)}{u_j(t)} / \alpha.
\end{eqnarray}
Analogously, it is called
\defined{$\alpha$-upper-dependent on the transition degree} if
\begin{eqnarray}
\max_{s \in T(A, m + 1)}{u_j(s)} \leq \alpha \max_{t \in T(A, m)}{u_j(t)}.
\end{eqnarray}
\end{defin}

This definition, together with Definition~\ref{def:bet_varied_util}, 
lets us prove (see appendix)
\begin{proposition}\label{prop:po_bound_limit_agentwise_cond}
Consider a game
$G = (N, S, (u_i)_{i = 1, \ldots, n})$
with a solution set $D \subseteq S$, such that
the utility of every player~$i$ is $\beta$ varied over $D$,
and it is $\alpha$-lower dependent on the transition degree over $D$ at transition degrees $i = 1, \ldots, m - 1$. Then,
\begin{eqnarray}
m-\pota \geq \poa / ((\prod_{i = 1}^{m - 1}{\alpha_i}) \beta).
\end{eqnarray}

If for every player~$i$, its utility 
is $\beta$ varied over $D$,
and $\alpha$-upper dependent on the transition degree over $D$ at transition degrees $i = 1, \ldots, m - 1$, then,
\begin{eqnarray}
m-\pots \leq (\prod_{i = 1}^{m - 1}{\alpha_i}) \beta \pos.
\end{eqnarray}
\end{proposition}

We exemplify the usage of this proposition in 
Example~\ref{ex:match_strat} in the appendix.

In Example~\ref{ex:match_strat}, Observation~\ref{obs:po_bound_limit}
provides a tight bound, which is,
in particular, tighter than the bound provided by
Proposition~\ref{prop:po_bound_limit_agentwise_cond}.
Observation~\ref{obs:po_bound_limit} always
provides a tight bound, given equality in its conditions. 
Still, Proposition~\ref{prop:po_bound_limit_agentwise_cond} can
be useful when its conditions are easier met.

%% file: ext_smooth.tex
\section{Efficiency of Transitions of Nash Equilibria}\label{Sec:ext_smooth}

This section specially treats the widely used case where the set of solutions is the set of 
Nash equilibria, denoted $\NE$. 
%
All the results from this section for (general) transitions keep holding if we talk about
limited transitions instead.

First, we observe
\begin{observation}\label{obs:ind_BR_TNE_eq_NE}
If the best response strategies of any player~$i$ to the
strategies $s_{-i}$ of the others do not depend on those $s_{-i}$,
then $\pota = \poa = \pos = \pots = 1$.
\end{observation}
\begin{proof}
Here, any transition of the
Nash equilibria is also a Nash equilibrium, since the best
responding still holds, so $T(\NE) = \NE$, implying the claim.
\end{proof}

\subsection{Constant-Sum and Potential Games}

We now deal with two broad and intuitively opposite classes of games: \emph{constant-sum} games,
where the interests of the players are opposed, and \emph{potential}
games, where there is a common goal, namely, the  potential function,
which is optimized. Constant-sum games
do not deteriorate from lack of coordination, since
the total utility is anyway zero, while potential games, where agents,
intuitively speaking, organize to maximize the potential function,
can suffer from lack of coordination.

We analyse these two kinds of games, and then 
decompose a general game into these kinds.
We immediately see that constant-sum games are insensitive to coordination.
\begin{observation}
In a constant-sum game, every profile has equal social welfare.
Thus, $\poa = \pos = \pots = \pota = 1$.
\end{observation}

In potential games, the social welfare can strongly deteriorate because
of a transition. We tightly bound such a
deterioration for \emph{congestion} games, which are
equivalent to the class of finite potential games%
~\cite[Theorems~{3.1, 3.2}]{MondererShapley1996PG}. We first remind the definition. 
\begin{defin}
A \defined{congestion} game~\cite{MondererShapley1996PG} is defined by a
set of agent $N = \set{1, \ldots, n}$, a set of resources
$M = \set{1, \ldots, m}$, a set of strategies $S = S_1 \times \ldots \times S_n$,
which are nonempty subsets of the resources,
i.e.~$S_i \subseteq 2^M \setminus{\emptyset}$, and cost functions
of the resources,
$\forall j \in M: c_j \colon \set{1, \ldots, n} \to \realsP$.

Next, for any strategy profile $s \in S$, let
$n_j(s) \defas \abs{i \in N : j \in s_i}$, and then we define the utility
of agent~$i$ as $u_i(s) \defas \sum_{j \in s_i}{c_j(n_j(s))}$,
finalizing the definition of the game.
\end{defin}

We state the next efficiency bound for subadditive cost functions.
Recall that
\begin{defin}\label{def:subadd}
A real-valued function $f$ is \defined{subadditive} if
$f(x + y) \leq f(x) + f(y)$.
\end{defin}

\begin{theorem}\label{Thm:cong_subadd_c_m_trans}
In a congestion game with subadditive cost functions,
$m-\pota \leq m \poa$, and this is tight. 
The bound holds for any solution concept, not
only \NE.
\end{theorem}
\begin{proof}
We shall prove an even stronger statement, which requires generalizing
the idea of transition to any profiles, not only solutions, as
follows.
\begin{defin}\label{def:merge}
Given a strategic form game
$G = (N, S = S_1 \times S_2 \times \ldots \times S_n, (u_i)_{i = 1, \ldots, n})$,
a \defined{merge} of strategy profiles $D \subset S$ is any transition
of the solution set $D$ (i.e.~when $D$ is taken as the solution set).
Similarly, the \defined{merge set} is defined to be $T(D)$.
\end{defin}
We now prove that for a congestion game with subadditive cost functions, the social welfare of
any merge is bounded by the sum of the total social welfare of the
constituent profiles.
\begin{lemma}\label{lemma:subadd_sw}
In a congestion game with subadditive cost functions, 
for any finite set of profiles $D$ 
and any merge $t \in T(D)$, there holds
$\sw(t) \leq \sum_{s \in D}{\sw(s)}$.
\end{lemma}
\begin{proof}
Recall that $n_j(s)$ is the number of agents using resource
$j$ in profile $s \in S$.
Fix $t \in T(D)$. There holds
\begin{eqnarray*}
\forall j \in M: n_j(t) \leq \sum_{s \in D}{n_j(s)}
\stackrel{\text{subadditivity}}{\Rightarrow} \forall j: c_j(n_j(t)) \leq \sum_{s \in D}{c_j(n_j(s))}\\
\Rightarrow \sw(t) \leq \sum_{s \in D}{\sw(s)}.
\end{eqnarray*}
\end{proof}

We now prove the theorem.

Lemma~\ref{lemma:subadd_sw} with any $m$ Nash equilibria (or other solutions)
implies the bound.

We demonstrate the tightness on the following example.
Consider an atomic routing game~\cite[\sectn{18.2.2}]{Nisanetal07} with $n$ parallel edges with $c(x) = x$ cost functions
between the source and the sink,
and assume that everyone of the $n$ players has to connect 
the source to the sink by an edge. Then, a profile is an equilibrium
if and only if the chosen edges are distinct.
Therefore, there exist $n!$ equilibria,
each of which costs $n$, so the price of anarchy is $1$. On the other hand, combining $m$
equilibria where various agents take the same route can
at worst make $m$ of the $n$ agents incur the cost of $m$ each,
thereby making the social cost become $m^2 + (n - m)$.
Thus, the price of $m$-transition anarchy is
$\frac{m^2 + (n - m)}{n}$, and for $m = n$, this becomes $n^2 / n = n$,
exactly $m = n$ times the price of anarchy.
\end{proof}

The utilities in a finite game where each player has the same number of
strategies can be decomposed to a zero-sum
game and a potential game~\cite{CandoganMenacheOzdaglarParrilo2011}.
The details are deferred to \sectnref{Sec:decomp_zero_sum_pot}. 

\subsection{Particular Cases}

We now study several neat bounds on the
efficiency of transitions.

\subsubsection{Games with Identical Utility Functions}


We prove the following where all the players
have the same utility, i.e.~$u_i \equiv u, \forall i \in N$.
\begin{proposition}\label{prop:id_util_eff}
For an identical utility game, 
the $\pos = \pots = 1$, but the price of anarchy can be arbitrarily low,
and the $\pota$ and even $\posta$ can be arbitrarily low relatively to the $\poa$.
\end{proposition}
\begin{proof}
Since the utilities are identical, any profile that maximizes the social
welfare maximizes also every player's welfare, thereby being an \NE.
Therefore, $\pos = 1$, and, by Observation~\ref{obs:po_bound_nat},
$\pots \geq \pos = 1$.
This part also follows from Proposition~\ref{prop:po_bound_agentwise_cond},
as Example~\ref{ex:pots_bound_agentwise_cond_id_util} demonstrates.
Actually, for two players, the equality of $\pots$ and $\pos$ also follows from
Proposition~\ref{prop:pots_pos_ne_2}.

As to the arbitrarily low price of anarchy, consider
the identical utility game described in
Matrix~\eqref{eq:id_util_game_low_poa}, for $0 < \epsilon < a$.
The set of Nash equilibria is $\set{(\epsilon, \epsilon), (a, a)}$, and
the price of anarchy can be made arbitrarily low by choosing small enough
$\epsilon$. As for the $\pota$ and $\posta$, they are simply zero here.
\begin{equation}
\begin{array}{l|c|c|}
		& 1:	& 2:	\\
\hline
I:	&	(\epsilon, \epsilon)	&  (0, 0)	\\
\hline
II:	& (0, 0)	&  (a, a)	\\
\hline
\end{array}
\label{eq:id_util_game_low_poa}
\end{equation}%
\end{proof}

\input{partic_case}

%% file: partic_case.tex

\subsubsection{Price of Stability}

For any solution concept,
to have 
$\pots \leq \alpha \pos$ ($\pota \geq \poa / \alpha$) means that for
any transition $t$ there exists a solution $s$ such that
$\sw(t) \leq \alpha \sw(s)$ ($\sw(t) \geq \sw(s) / \alpha$).
The bounding solution can happen to be one of the solutions composing the transition.
Namely, if for each transition $t = (s_1, \ldots, s_n) \in T(\NE)$, there
exists $i \in N$ and $(s_i, d_{-i}) \in \NE$, such that 
$\sw(s) \leq \alpha \sw((s_i, d_{-i}))$, then $\pots \leq \alpha \pos$.
By analogy, always having $i \in N$ and $(s_i, d_{-i}) \in \NE$
such that $\sw(s) \geq \sw((s_i, d_{-i})) / \alpha$ would imply
$\pota \geq \poa / \alpha$.
In particular,
\begin{proposition}\label{prop:pots_pos_ne_2}
In a two-player game, if for every $x, x' \in S_1$
there holds the implication
\begin{eqnarray}
u_1(x, y) \leq u_1(x', y) \text{ and } u_2(x, y) \leq u_2(x, y') \nonumber\\
\Rightarrow
\sw(x, y) \leq \sw(x', y) \text{ or } \sw(x, y) \leq \sw(x, y'),
\label{eq:util_imply_sw}
\end{eqnarray}
then we have $\pots = \pos$.
\end{proposition}
The proof uses the definitions of transitions and \NE{}.

\begin{proof}
Take any transition $(x, y) \in T(\NE)$. By the definition of a
transition, there exist $x' \in S_1$ and $y' \in S_2$ such that
$(x', y)$ and $(x, y')$ are Nash equilibria. By the definition of
an \NE, $u_1(x', y) \geq u_1(x, y)$ and $u_2(x, y') \geq u_2(x, y)$.
Now, \eqnsref{eq:util_imply_sw} implies that 
$\sw(x, y) \leq \sw(x', y) \text{ or } \sw(x, y) \leq \sw(x, y')$,
which means that $\sw(x, y)$ is upper bounded by the social welfare of
at least one Nash equilibrium. Since $(x, y)$ is any transition, this
implies that $\pots \leq \pos$. Together with the opposite inequality,
given in Observation~\ref{obs:po_bound_nat}, we obtain $\pots = \pos$. 
\end{proof}

\subsubsection{Extensive Smoothness}

We now extend the notion of a smooth game used to bound the price of
anarchy~\cite{Roughgarden2015} to be able to bound the price of transition
anarchy. We do not attempt to allow for the smoothness to be generalizable,
because the notion transition allows only for pure profiles. We can take
the solution set to be mixed or, say, correlated Nash equilibria, but
the transitions themselves are not mixed, because they are intended to
represent choices of strategies from different solutions by the players.

Since the notion of a transition,
allowing arbitrarily picking strategies from various Nash equilibria,
is much broader than the notion of an \NE, 
connecting the social welfare in a transition
with the optimal possible social welfare is harder,
requiring
the following notion of \defined{extensive smoothness}.

\begin{defin}
A game\footnote{For consistency within the paper, we
assume a utility maximization game, unlike~\cite{Roughgarden2015} does.}
is \defined{$\alpha, \beta, \lambda, \mu$-extensively smooth} if the
following holds:
\begin{enumerate}\label{ext_smooth}
	\item \label{ext_smooth:obt_ne_close}
	$\forall i \in N, \forall s \in T(\NE), \forall d \in \NE \text{ such that } s_i = d_i:
	u_i(s) \geq \alpha u_i(d)$.
	This means that substituting the strategies of the agents
	besides $i$ so as to obtain an \NE{} from the transition
	does not increase $i$'s utility much.
	
	\item	\label{ext_smooth:trans_close}
	$\forall s^* \in \arg\max\set{\sw(s) : s \in S}, \forall t, v \in T(\NE):
	u_i(s^*_i, t_{-i}) \geq \beta u_i(s^*_i, v_{-i})$.
	This requires that the concrete transition that completes
	a strategy from a socially optimum profile does not matter much to $i$'s
	utility.
	
	\item	\label{ext_smooth:orig_smooth}
	$\forall s^* \in \arg\max\set{\sw(s) : s \in S}, \forall t \in T(\NE):
	\sum_{i = 1}^n {u_i(s^*_i, t_{-i})} \geq \lambda \sw(s^*) - \mu \sw(t)$.
	This is the generalization of the original $\lambda, \mu$-smoothness,
	where the original assumption $t \in \NE$, which is enough if we do not intend
	to generalize the bound beyond the pure case, is generalized to
	$t \in T(\NE)$.
\end{enumerate}
\end{defin}

This allows to bound the price of transition anarchy.
\begin{proposition}\label{prop:ext_smooth:pota}
Any $\alpha, \beta, \lambda, \mu$-extensively smooth game has
$\pota \geq \frac{\alpha \beta \lambda}{1 + \alpha \beta \mu}$.
\end{proposition}
\begin{proof}
For any transition $s$ and socially optimal $s^*$, 
\begin{eqnarray*}
\sw(s)
\geq \alpha \sum_{i \in N}{u_i(s_i, d_{-i}(i))}
\geq \alpha \sum_{i \in N}{u_i(s^*_i, d_{-i}(i))}\\
\geq \alpha \beta \sum_{i \in N}{u_i(s^*_i, s_{-i}(i))}
\geq \alpha \beta (\lambda \sw(s^*) - \mu \sw(s)).\\
\Rightarrow \sw(s) \geq \frac{\alpha \beta \lambda}{1 + \alpha \beta \mu} \sw(s^*)
\iff \pota \geq \frac{\alpha \beta \lambda}{1 + \alpha \beta \mu}.
\end{eqnarray*}
The first inequality stems from
condition~\ref{ext_smooth:obt_ne_close} of the proposition,
if we choose $d_{-i}(i)$ so as to complete $s_i$ to a Nash equilibrium.
The second inequality follows from the definition of an \NE,
the third one stems from condition~\ref{ext_smooth:trans_close},
and the fourth inequality follows from
condition~\ref{ext_smooth:orig_smooth}.
\end{proof}

%% file: eff_class_game.tex
\section{Routing Games}\label{Sec:rout_game}

We now provide tight bounds for the efficiency of transitions in routing games,
which are not usual games with explicit players, and therefore, require a
special treatment.


Consider the famous routing games, modeling large
traffic where every driver controls an infinitesimally tiny part of the whole load~\cite{Wardrop1952}.
In a \defined{non-atomic routing game}~\cite[\sectn{18.2.1}]{Nisanetal07}
or simply a \defined{routing game} on network $G = (V, E)$ with $k$ source
and sink pairs $(s_1, t_1), \ldots, (s_k, t_k)$, each \defined{commodity~$i$} of size $r_i$ 
is to be routed from $s_i$ to $t_i$ through
the paths in $\calP_i$. Define $\calP \defas \cup_{i = 1}^k {\calP_i}$.
A \defined{flow} vector $f \in \realsP^{\abs{\calP}}$
is \defined{feasible} if $\sum_{P \in \calP_i} {f_P} = r_i$
for each $i = 1, \ldots, k$. For each edge~$e$ we are given a
non-decreasing \defined{cost function $c_e \colon \realsP \to \realsP$},
and we define the cost of a path $P$ as
\begin{equation}
c_P(f) \defas \sum_{e \in P}{c_e(f_e)},
\label{eq:nonat_cost_path}
\end{equation}
where $f_e \defas \sum_{P \in \calP : e \in P}{f_P}$.
Finally, define an \defined{equilibrium flow} as a feasible flow~$f$
such that for every commodity $i = 1, \ldots, k$, for every path
$P \in \calP_i$ such that $f_P > 0$ and for every path $P' \in \calP_i$
we have
\begin{equation}
c_P(f) \leq c_{P'}(f).
\label{eq:nonat_eq_flow}
\end{equation}

Define \defined{the cost of flow~$f$} as the averaged cost of the various
paths $P \in \calP$; formally,
\begin{eqnarray}
C(f) \defas \sum_{P \in \calP}{c_P(f) \cdot f_P} = \sum_{e \in E}{c_e(f_e)} \cdot f_e,
\label{eq:nonat_cost_flow}
\end{eqnarray}
the second version of the definition being obtained by substituting the
definition of $c_P(f)$ (\eqnsref{eq:nonat_cost_path}) and changing the
summation order.

Define \defined{the price of anarchy and
stability} as the ratio of the cost of an equilibrium flow (the same cost
for every equilibrium flow~\cite[Theorem~{18.8}]{Nisanetal07}) to the
cost of the cheapest possible feasible flow.

We now define
a \defined{transition} of equilibrium flows analogous to how we defined
the transition of a solution concept, i.e., by allowing for an arbitrary
playing of various solutions. Here, however, a single
decision maker controls an infinitesimal part of the flow, and so a player
playing a given equilibrium flow translates to, generally speaking,
playing possibilities on each of the paths where the flow is
positive. Therefore, combining such plays arbitrarily can result in
arbitrary reallocation of the flow of each commodity, resulting,
sometimes, in infinitely many transition flows even for a single equilibrium flow.
Formally, a \defined{transition} is a feasible flow $f$ such that for
every commodity~$i$ and path $P \in \calP_i$ with $f_P > 0$, there exists
an equilibrium flow $f'$, such that $f'_P > 0$.
Intuitively, we need an equilibrium flow to be positive to say that
every infinitesimal player of commodity~$i$ flows through path
$P$ in some equilibrium, allowing us to have any flow on this path in
a transition.
\footnote{Applying the idea of stable transition to a non-atomic
routing game yields the same concept as equilibrium flow, since
a single player's deviation does not matter here.}

This lets us define \defined{the price of transition anarchy (stability)
} as the ratios of the costs of the most (least) costly
transition to the cheapest feasible flow.

First, we exemplify that a game with a
unique equilibrium flow can have infinitely many transitions because of
the arbitrary leeway in reshaping the flow in a transition.
\begin{example}\label{ex:eq_1_trans_inf}
Consider the network in \figref{fig:n_par_x}, where the only commodity
has size $r = 1$. The only equilibrium flow has $1 / n$ over each edge
and costs $1 / n$.
This game admits a continuum of transitions: every
flow $f$ such that $\sum_{P \in \calP}{f_P} = 1$ is a transition.

Regarding efficiency, since the only equilibrium flow is also
socially optimal, $\poa = \pos = \pots = 1$. The transitions, however, can let all
the commodity flow through a single edge, thereby incurring the cost to
$1$. Therefore, $\pota = n$.

\begin{figure}[htb]
\center
\includegraphics[trim = 0mm 0mm 0mm 0mm, clip=true,width=0.31\textwidth]{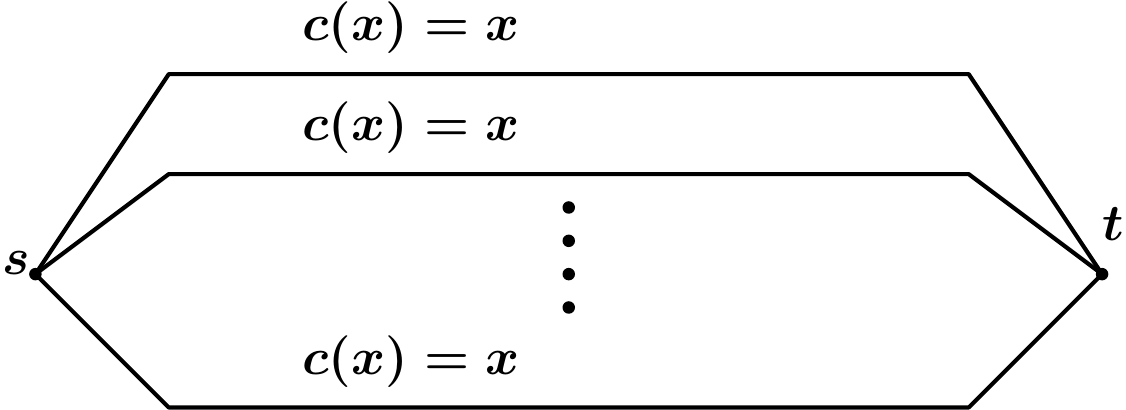}
\caption{Having $n$ parallel edges with $c_e(x) = x$ each.
}%
\label{fig:n_par_x}
\end{figure}
\end{example}

In the last example we witnessed a remarkable phenomenon of a single
equilibrium flow yielding a continuum of transitions and the price of
transition anarchy being $n$ times the price of anarchy. The natural
next question is to which extent transitions can have more 
costs than equilibrium flows. The following theorem resolves this by
tightly bounding the price of anarchy of a routing game.

\begin{theorem}\label{the:poa_up_bound_tight}
Given a set of cost functions $\calC$, a routing game and a
commodity~$i$, define 
\begin{equation}
S_i(\calC) \defas
{ \frac{ \max\set{\abs{P} : P \in \calP_i} \sup_{c \in C}{(c(r_i + \sum_{j \in \set{1, \ldots, k} \setminus \set{i}}{r_j}))} }
{ \min\set{\abs{P} : P \in \calP_i} \inf_{c \in C}{c(r_i / \abs{\calP_i})} } }.
\label{eq:stretch}
\end{equation}

Then, any routing game with cost functions $\calC$ has
$\pota \leq \poa \cdot \max_{i = 1, \ldots, k}{S_i(\calC)}$,
and this bound is tight, in the sense that there exist games,
such that their $\pota$ is arbitrarily close to
$\poa \cdot \max_{i = 1, \ldots, k}{S_i(\calC)}$.

In particular, for linear cost functions $c_e(x) = a_e \cdot x$,
if we denote by $a_{\max}$ and $a_{\min}$ the largest and the smallest
coefficients, respectively, we obtain
\begin{equation}
S_i(\calC) = { \frac{ \max\set{\abs{P} : P \in \calP_i} {(a_{\max} (r_i + \sum_{j \in \set{1, \ldots, k} \setminus \set{i}}{r_j}))} }
{ \min\set{\abs{P} : P \in \calP_i} a_{\min}{r_i} } } \abs{\calP_i}.
\label{eq:stretch_lin}
\end{equation}
If we additionally assume that the paths of different commodities never
intersect, then we can dispose of the
$\sum_{j \in \set{1, \ldots, k} \setminus \set{i}}{r_j}$ term, ending up
with
\begin{equation}
S_i(\calC) = { \frac{ \max\set{\abs{P} : P \in \calP_i} {a_{\max}} }
{ \min\set{\abs{P} : P \in \calP_i} a_{\min} } } \abs{\calP_i}.
\label{eq:stretch_lin_no_intersect}
\end{equation}
\end{theorem}

The proven tight bound is, indeed, quite appalling, admonishing
coordination, but it does get optimistic in some cases. For instance,
for linear costs, not intersecting commodities and similar path lengths
withing the paths for each given commodity,
\eqnsref{eq:stretch_lin_no_intersect} becomes approximately
$\frac{a_{\max}}{a_{\min}} \abs{\calP_i}$. Therefore, in such cases and for comparable
cost functions and few paths for a commodity, the blowup caused
by the lack of coordination is small.

As to the price of transition stability, it can either be 
equal to the price of stability, strictly between the $\pos$ and $1$
or equal to $1$. Examples are omitted due to lack of space.

For networks with parallel paths,
we now 
provide a condition when $\pots = 1$, which is proven in the appendix.
\begin{proposition}\label{prop:par_path_pots}
Let the network have edge-disconnected paths, i.e.~no two paths
$P, P' \in \calP$ share edges. Then, the flow through all the edges of
a path is the same, and we can see the cost functions of paths are real
functions, $c_P \colon \realsP \to \realsP$.

Assume that the cost functions $c_e$ are continuous
and (strictly) increasing. 
If for each commodity $i$, $c_P(0) = c_{P'}(0)$ for every
$P, P' \in \calP_i$, then $\pots = 1$.
\end{proposition}

We can naturally generalize the definition of an $m$-transition for flows,
by allowing a positive flow if there exists a set of $m$ equilibrium flows
at least one of which is positive on the path.
Example~\ref{ex:eq_1_trans_inf} holds, demonstrating that a unique
equilibrium flow can allow for infinitely many $m$-transitions, for every
$m \geq 1$. The bound of Theorem~\ref{the:poa_up_bound_tight}
remains tight, since the tight example uses a single equilibrium flow.

%% file: stable_trans_coord_game.tex
\section{Stable Transitions and Coordination Games}\label{Sec:stable_trans_coord_game}

We now analyse stable transitions, concentrating
on coordination games~\cite{AptdeKeijzerRahnSchaeferSimon2017}.
These basic games model the essence of coordination. 
We assume that all the nodes have two possible colours $\set{1, 2}$,
unless stated otherwise.
We need to restrict transitions, since they
allow for all the possible combinations of $1$ and $2$, because
everyone choosing $1$ is an equilibrium, and so is everyone choosing $2$.

We first formally define coordination games,
following~\cite{AptdeKeijzerRahnSchaeferSimon2017}.
\begin{defin}
A \defined{coordination game} on undirected graph $G = (N, E)$
has players $N$. Each player~$i$ is given a set of colours $S_i$,
which constitute the player's strategy set.\footnote{As mentioned,
our default is that $S_1 = \ldots = S_n = \set{1, 2}$.}
The utility of player~$i$
is the number of her neighbours that pick the same colour, i.e.~%
$u_i(s) \defas \abs{\set{j \in \Neighb(i) : s_i = s_j}}$, representing
the coordination of $i$ with its neighbours.
\end{defin}

Every combination of the
colours $1$ and $2$ constitutes a transition, while stable transitions
are much more restrictive. For example, in the case of the star topology
with $n - 1$ leaves, the only stable transitions that are not \NE{}
colour the cetre in $1$ and 
$\floor{\frac{n - 1}{2}}$ leaves in $1$ and the rest $\floor{\frac{n}{2}}$
leaves in $2$, or fully switching between $1$ and $2$. We 
restrict the stable transitions and \NE{} in the following observation.
We note that Theorem~$8$
in~\cite{AptdeKeijzerRahnSchaeferSimon2017} uses different strategy sets,
while we assume the set $\set{1, 2}$ for everyone.
\begin{observation}\label{obs:coord_game:low_sw_bound}
Consider a coordination game, where
profile $s$ is a stable transition. Then, every player~$i$ has
at least $\floor{\frac{\deg(i) - 1}{2}}$ neighbours of its colour.
In an equilibrium, every player has at least
$\ceil{\frac{\deg(i)}{2}}$ neighbours of its colour.
\end{observation}
Thus, to check whether a given profile is a stable transition or not, do this:
\begin{enumerate}
\item	For each player $i$,
	\begin{enumerate}
		\item If $i$ has less then $\floor{\frac{\deg(i) - 1}{2}}$ neighbours of its colour,
		return ``No''.
		\item	Else, if $i$ has exactly $\floor{\frac{\deg(i) - 1}{2}}$ neighbours of its colour,
		consider $j \in \Neighb(i)$ with less than $\ceil{\frac{\deg(i)}{2}}$ neighbours of its colour.
		\item	If no such $j$ exists, return ``No''.
	\end{enumerate}
\item	If we have not returned by now, return ``Yes''.
\end{enumerate}

We use the observation to prove the following efficiency theorem.
\begin{theorem}\label{The:eff_low_bound}
In a coordination game,
$\posta \geq \frac{1}{2} - \frac{\abs{N}}{2 \abs{E}}$,
$\poa \geq 1 / 2$,
and these bounds are tight.
\end{theorem}
In particular, the denser the graph is, the closer the lower bound
for the price of stable transition anarchy becomes to $1 / 2$.

Since $D \subseteq ST(D)$, $D \neq \emptyset \iff ST(D) \neq \emptyset$. The interesting
existence question, addressed by the following theorem, is when
there exist stable transitions that are not equilibria, i.e.~%
$ST(\NE) \setminus \NE \neq \emptyset$.
\begin{theorem}\label{The:st_exist}
\begin{enumerate}\label{enum:st_exist}
    \item   \label{enum:st_exist:cycle}
		Any cycle of length at least $4$ has a stable transition
    that is not an \NE.
		For an \emph{even} $n$, this can actually attain the lower
		bound from Theorem~\ref{The:eff_low_bound}.
    \item   \label{enum:st_exist:clique}
		For a complete graph on $n$ nodes, there holds
    $ST(\NE) \setminus \NE \neq \emptyset \iff n$ is \emph{even}.
    \item   \label{enum:st_exist:forest}
		For a forest that contains not only isolated nodes, there holds
    $ST(\NE) \setminus \NE \neq \emptyset$.    
\end{enumerate}
\end{theorem}

This theorem suggests how to convert a given graph to a graph where there
will be no stable transitions besides the equilibria. For instance,
given a clique with an even number of nodes, we can remove one node
and obtain an odd clique, where only equilibria constitute 
stable transitions.

%% file: alg.tex
\section{Algorithms and Complexity}\label{Sec:alg_comp}

We now tackle algorithmic issues, assuming we are given the 
set of solutions as a list. First, 
deciding whether a given profile constitutes a transition
amounts to checking for each player, whether there exists 
a solution where this player acts as it does in the given profile.
Deciding about constituting a stable transition for coordination
games was done in \sectnref{Sec:stable_trans_coord_game}.

Enumerating all the transitions or all the $m$-transitions
is trivial from definition.
Enumerating all the stable transitions in a coordination game
can be done by checking all the profiles exhaustively, for each
profile using the algorithm from \sectnref{Sec:stable_trans_coord_game}.

Consider the questions of finding 
the least $m$ such that a given transition is an $m$-transition,
and deciding, given a game, whether the set 
$T(D) \setminus T(D, m)$ is
nonempty for a given $m$ (observe that for any $m \geq 2$,
$T(D, m) \setminus D \neq \emptyset \iff T(D) \setminus D \neq \emptyset \iff$
$D$ is not a product of sets).

\subsection{The minimum $m$ such that $t \in T(D, m)$}

We prove that this problem is approximation equivalent to
the famous \defined{set cover}~\cite{Karp1972}.
In \defined{set cover}, we are given $(S_i)_{i = 1}^m$
and need to find a smallest subcollection $((S_{i_j})_{j \in J})$
such that $\cup_{j \in J}{S_{i_j}} = \cup_{i = 1}^m {S_i}$.
This immediately implies a 
$1 + \ln(n)$-approximation~\cite{Johnson1974} and 
inapproximability within $c \log(n)$, for some $c > 0$%
~\cite{RazSafra1997}.
\begin{theorem}
Given a profile $t \in T(D)$, finding the smallest $m$ 
such that $t \in T(D, m)$ is approximation-equivalent to
set cover.
\end{theorem}
\begin{proof}
The reduction to set cover works as follows. First,
given a the list of solutions $D$, we say solution~$s \in D$ 
\defined{covers} player~$i$'s strategy in $t$ if $s_i = t_i$.
Then, we eliminate all the solutions covering no strategies
in $t$, so that each solution will cover some strategies.
Now, considering each solution $s \in D$ as a set and the strategies 
$(t_i)_{i = 1}^n$ as elements, we reduce our problem to covering 
the set $\set{t_i : i \in N}$ using the minimum number of sets $s \in D$.

Next, we reduce set cover to our problem.
Given a set cover instance $(S_i)_{i = 1}^m$, let the set of
elements be $U \defas \cup_{i = 1}^m {S_i}$. We define a game on
$\abs{U}$ players, each player~$i$ having $S_i \defas \set{0, 1}$
and let $t$ be $(1, \ldots, 1)$, a vector of length $\abs{U}$.
For each set $S_i$, we define a vector $d(i)$ as follows:
$$d(i)_j \defas
 \begin{cases}
    1, & \text{if } j \in S_i, \\
    0, & \text{otherwise.}  \\
  \end{cases}
$$
Thus, $t_j = d(i)_j \iff j \in S_j$.
Now, covering $U$ with sets $\calC \subseteq (S_i)_{i = 1}^m$ is 
equivalent to obtaining $t$ from the profiles
$\set{d(i) : S_i \in \calC}$, which we define to be the solution set $D$.
Therefore, finding the minimum $m$ such that $t \in T(\calC, m)$
solves the given instance of set cover.
\end{proof}

\subsection{Strictness of Inclusion}

We now tackle the question about $T(D, m)$ being a proper subset of $T(D)$ for a given $m \geq 2$. 
In other words, whether we can attain any transition using at most $m$ different solutions.
%
We now prove that deciding whether $T(D) = T(D, m)$ is simple.
\begin{theorem}
\begin{enumerate}
	\item Any maximal set solutions such that no body is 
	in a transition of the others is of the same size.
	\item	Therefore, we can arbitrarily add solutions till
	everyone is a transition of the created set. The size
	of that set is the minimum $m$ satisfying $T(D) = T(D, m)$.
\end{enumerate}
\end{theorem}
\begin{proof}
First, we observe that sets of solutions such that nobody
constitutes a transition of the others constitute a matroid,
where the independent sets are those where nobody is a transition
of the others.
Therefore, the maximal such sets are of the same size, being
the bases of the matroid. Therefore, arbitrary addition
suffices.
\end{proof}

Being a matroid allows us also execute the algorithms which 
are available for matroids.

%% file: background.tex
\section{Background}\label{Sec:background}

The most famous solution set to a game is, arguably,
Nash equilibria~\cite{Nash51}, though many other concepts, such as, for
example, strong \NE~\cite{Aumann59}
and approximate equilibria~\cite{GovindanWilson05}, are widely used.
There is much interest in the epistemic underpinnings of
interaction~\cite{NehringBonanno03,Perea2012}, and here we take another
approach to modeling interaction.

The above mentioned central efficiency measures, \defined{price of anarchy} and
\defined{price of stability}, express~\cite[\sectn{17.1.3}]{Nisanetal07} 
the worst possible stable situation independent agents can have and
the best possible stable situation we can suggest to the independent
agents, respectively. When applied to transitions, we look at the ratio
of the socially worst or the best transition to the social optimum.
The socially worst transition represents the worst possible situation
independent agent that are not coordinated on a solution can have,
while the socially best transition cannot be described as a profile
to suggest, because it is not necessarily a solution. This is still interesting
as the best option of what agents that are not coordinated can
have.

The notion of efficiency of transition considers non-coordinated play of
equilibria, while there are studies of (coordinated) equilibria
played by biased agents~\cite{MeirParkes2015}.

Constant-sum games constitute a widely known and studied class
of games~\cite{vonNeumannMorgenstern1944} and so do potential games.
Potential functions were first used for congestion games by
Rosenthal~\cite{Rosenthal1973}, games which were proven to be isomorphic
to all potential games in~\cite{MondererShapley1996PG}.

The routing games studied in \sectnref{Sec:rout_game} were partially
discussed by Pigou~\cite{Pigou1932}, but
Wardrop~\cite{Wardrop1952} was the first to define them formally.
Efficiency of routing is of paramount importance, being the motivation
and a test for many techniques, brilliantly described in~\cite{Roughgarden2005}. Predicting
traffic and its efficiency is needed~\cite{Ozdaglar2008,Ozdaglar2009},
and we tightly bound the cost of incoordination.

Existence, efficiency and complexity of $k$-equilibria of coordination
graphical games were studied
in~\cite{RahnSchaefer2015,AptdeKeijzerRahnSchaeferSimon2017},
whereas the general graphical games were defined by
Kearns et al.~\cite{KearnsLittmanSingh2001},
and the polymatrix games were first studied
in~\cite{Janovskaya1968,Howson1972}.
Various network games are well covered in~\cite{JacksonZenou2015}.

%% file: conclusion.tex
\section{Conclusions and Future Work}\label{Sec:conclusion}

In order to model uncoordinated transitions between solutions
and estimate their efficiency,
we define the notion of a transition and its efficiency measures, namely
the price of transition anarchy and transition stability.

While we show that in general games, efficiency bounds on the
solutions, such as Nash equilibria, do not guarantee much regarding the
efficiency of the transitions, we do bound the efficiency of the transitions
by a factor of the efficiency of the solutions,
provided we have bounds on the changes in the utilities because of
the change from a solution to a transition. If the solutions are
Nash equilibria, we fully analyse the basic games of opposed
interests - constant-sum games, and the games of aligned
interests - congestion games with subadditive costs. These
two classes of games allow bounding the efficiency loss that
general games with equally sized strategy sets for all agents
incur.
We also generalize the smoothness condition to bound
the worst efficiency a transition can have.
For two players, given a connection between the individual
utilities and the social welfare, we prove that no transition
can be more efficient than all the \NE. 
For identical utility games, we provide several optimality results
for efficiency. For routing games, we show that even when all the 
equilibrium flows are socially optimal, the game can still possess
very inefficient transitions. We provide a tight bound on the worst
efficiency of a transition, which indicates that for linear costs,
not intersecting commodities, similar path lengths for each commodity,
and few paths per commodity,
the blowup is approximately as large as the ratio of the coefficients of
the cost functions. On the other hand, we also show when the optimal
transition can bring to the socially optimal profile.
We  also provide tight efficiency bounds on transitions and more
restricted stable transitions of coordination games.
To summarize, try to avoid uncoordinated transitions
between solutions, unless the game is of one of the types we have proven to
preserve efficiency even in transitions.
If we can assume the restricted stable transitions, then
we provide optimistic efficiency bounds and suggestions how to
change the graph of the coordination game to avoid being
non-coordinated.

Curiously, the computation of the smallest transition
degree to render a given profile a transition is \NP-hard, while
the computation of the smallest transition degree making all 
the profiles transitions is possible, due to the matroid
structure.

Many interesting future directions exist.
First, since transitions generally ruin efficiency unboundedly, 
we consider two major constraints on transitions, namely limited
transitions and stable transitions. The former limit which
solutions constitute a transition, while the latter rule out
unreasonable transitions. More similar constraints may be 
considered in an attempt to model reality and derive further
efficiency bounds.
Another important direction is transitions of extensive
games. The subgame perfect equilibria will simply keep holding,
while various equilibria for imperfect information games 
require a deep treatment.

Additionally, we would like to replace all the efficiency
bounds with tight results and obtain a full characterisation
when the stable transitions include only the equilibria. One can specifically study concrete games, 
like we did for routing games and coordination games. 
In the latter games, relaxing the assumption of two colours $\set{1, 2}$ 
per player is the next step.
As to the applicability of the concept of transition
and limited and stable transition,
empirical work is required to examine how people, companies and
other agents transit from one equilibrium to another one. 
Algorithmic questions, such as checking whether a given profile is
an $m$-transition for a given $m$ and which transition is the least socially efficient, are theoretically and practically interesting.
Modeling rational choosing between the various solution as a repeated
game between communicating agents is an important goal.
%
\input{extensions}

In summary, we model transitions between solutions
and thoroughly analyse the efficiency of such transitions.

%% file: extensions.tex


To express that even a single agent hesitates between 
several solutions and combines several such solutions, we can also define
\defined{combined transitions} and generalize some of the results of the
paper to them.

We can combine solutions from \emph{different} solution concepts;
for instance, combining the rational \NE{} for prisoner's dilemma
(both defect) with the superrational solution~\cite{Hofstader1983}
(both cooperate) fits the experimental results.

%% file: omit_proofs.tex
\section{Omitted Proofs and Examples}\label{Sec:omit_proofs_etc}

We first prove Observation~\ref{obs:po_bound_nat}.

\begin{proof}
As to the price of anarchy, there holds
\begin{equation*}
\pota = \frac{\min_{s \in T(D)}{\sw(s)}}{\max_{s \in S}{\sw(s)}}
\stackrel{D \subseteq T(D)}{\leq} \frac{\min_{s \in D}{\sw(s)}}{\max_{s \in S}{\sw(s)}}
= \poa,
\end{equation*}
where the first equality follows from
Definition~\ref{def:efficiency_measures}, and the second equality does from the
definition of the price of anarchy.

For the price of stability, we similarly have
\begin{equation*}
\pots = \frac{\max_{s \in T(D)}{\sw(s)}}{\max_{s \in S}{\sw(s)}}
\stackrel{D \subseteq T(D)}{\geq} \frac{\max_{s \in D}{\sw(s)}}{\max_{s \in S}{\sw(s)}}.
\end{equation*}
\end{proof}

We show that 
without further assumptions, 
the price of anarchy can strongly differ from the price of transition
anarchy. 
\begin{example}\label{ex:pota_can_dif_poa}
In the coordination game of Matrix~\eqref{eq:coord_game_ineff}, both
Nash equilibria are socially optimal, while any mix of them which is not
an equilibrium achieves zero utility for both agents. Thus,
$\poa = \pos = 1$, while $\pota = 0$.
\end{example}

The next example demonstrates the possible gap between the price of
stability and the price of transition stability.
\begin{example}\label{ex:pots_can_dif_pos}
Consider the following game: players $1$, $2$ and $3$ choose
between the strategies $0$ and $1$ each. For each outcome besides
$(0, 0, 0)$, all the players' utilities are $b > 0$. When the outcome
is $(0, 0, 0)$, the utility of player~$1$ is $a$, for $a \gg b$, and
the others receive zero. In this game, all the strategy profiles
besides $(0, 0, 0)$ and $(1, 0, 0)$ are Nash equilibria, because from any
such profile the only way the utility of any player changes is that a
player other than $1$ deviates, whereby obtaining not more than it already
obtains. At profile $(1, 0, 0)$ player $1$ can move to $(0, 0, 0)$,
thereby increasing its utility, and at $(0, 0, 0)$ player $2$ or $3$ can
unilaterally increase its own utility to $b$.
Therefore, $\pos = \frac{3b}{a}$.
However, any strategy of any player appears in some \NE, and therefore,
the transition set contains all the  possible profiles. Thus, $\pots = 1$.
\end{example}

We now prove Observation~\ref{obs:po_bound}.

\begin{proof}
If $\min_{s \in T(A)}{\sw(s)} \geq \min_{t \in A}{\sw(t)} / \alpha$, then
\begin{eqnarray*}
\pota = \frac{\min_{s \in T(D)}{\sw(s)}}{\max_{s \in S}{\sw(s)}}
\geq \frac{\min_{s \in D}{\sw(s)} / \alpha}{\max_{s \in S}{\sw(s)}}
= \poa / \alpha,
\end{eqnarray*}
where the first equality follows from
Definition~\ref{def:efficiency_measures}, the first inequality follows
from the assumption,
and the final equality
uses the definition of the price of anarchy.
We also observe that the only inequality becomes equality if and only if
the assumption inequality holds with equality. 

The proof of \eqnsref{eq:po_bound:s:up} is analogous.
\end{proof}

We present an example of using Observation~\ref{obs:po_bound}.
\newcounter{ex_coord_game_alpha}
\setcounter{ex_coord_game_alpha}{\value{example}}
\begin{example}\label{ex:coord_game_alpha}
In the coordination game in Matrix~\eqref{eq:coord_game_alpha},
let $a > b > a / c$. There
exist two \NE: $(I, 1)$, because $a > b$,
and $(II, 2)$, because $b > a / c$. 
The maximum possible social welfare is $2 a$, because $a > b$.
Therefore, $\poa = b / a$ and $\pos = 1$.

$T(\NE) = S$, 
which means that
$\min_{s \in T(\NE)}{\sw(s)} = \min_{t \in \NE}{\sw(t)} / \alpha$,
for $\alpha = \frac{2 c b}{a + b}$ and 
$\max_{s \in T(A)}{\sw(s)} = 1 \cdot \max_{t \in A}{\sw(t)}$,
Therefore, Observation~\ref{obs:po_bound}
implies that $\pota = \poa / \left(\frac{2 c b}{a + b}\right)
= \frac{a + b}{2 c a}$ and $\pots = 1 \cdot \pos = 1$.
\end{example}

\begin{equation}%
\begin{array}{l|c|c|}
		& 1:	& 2:	\\
\hline
I:	&	(a, a)	&  (a / c, b / c)	\\
\hline
II:	& (b / c, a / c)	&  (b, b)	\\
\hline
\end{array}
\label{eq:coord_game_alpha}
\end{equation}

We next prove Proposition~\ref{prop:po_bound_agentwise_cond}.

\begin{proof}
We begin with $\poa$ and $\pota$. 
If player $i$'s utility over $D$ is $\alpha$-lower dependent
on coordination, then for every transition $s \in T(D)$,
we have $u_i(s) \geq u_i(s') / \alpha$, where $s'$ is some solution.
Let $t$ be a solution with the smallest possible
social welfare (perhaps, $s' = t$). This implies, by the condition of
the proposition, that $u_i(s') \geq u_i(t) / \beta$, implying that
$u_i(s) \geq u_i(t) / (\alpha \beta)$. Therefore, 
\begin{eqnarray*}
\pota = \frac{\min_{s \in T(D)}{\sw(s)}}{\max_{s \in S}{\sw(s)}}
= \frac{\min_{s \in T(D)}{\sum_{i \in N}{u_i(s)}}}{\max_{s \in S}{\sw(s)}}\\
\geq \frac{{\sum_{i \in N}{u_i(t) / (\alpha \beta)}}}{\max_{s \in S}{\sw(s)}}
= \frac{\min_{t \in D}{\sw(t)} / (\alpha \beta)}{\max_{s \in S}{\sw(s)}}
= \poa / (\alpha \beta).
\end{eqnarray*}

We can prove \eqnsref{eq:po_bound_agentwise_cond:s_up} analogously.
\end{proof}

We next give two examples of using Proposition~\ref{prop:po_bound_agentwise_cond}.
\begin{example}[Identical utility games]\label{ex:pots_bound_agentwise_cond_id_util}
For games with identical utility functions,
a profile that maximizes the social welfare maximizes also everyone's
utility, thus being an \NE. Therefore, such a game is $1$-upper
dependent on coordination. It is also $1$ varied over any set,
and Proposition~\ref{prop:po_bound_agentwise_cond} implies that
$\pots \leq \pos$, and therefore, $\pots = \pos = 1$.
\end{example}

We exemplify this proposition by continuing
Example~\ref{ex:coord_game_alpha}.
\newcounter{tmp}
\setcounter{tmp}{\value{example}}
\setcounter{example}{\value{ex_coord_game_alpha}}
\begin{example}[Cont.]
The utility of every player over the set of Nash equilibria is
$c$-lower and $1$-upper dependent on coordination.
In addition, every agent's utility is $1$ varied
over $\NE$. Therefore, Proposition~\ref{prop:po_bound_agentwise_cond}
implies that $\pota \geq \poa / c$ and $\pots = 1 \cdot \pos$.
\end{example}
\setcounter{example}{\value{tmp}}

In Example~\ref{ex:coord_game_alpha}, Observation~\ref{obs:po_bound}
provides a tight bound, while 
Proposition~\ref{prop:po_bound_agentwise_cond} does not.
Actually, Observation~\ref{obs:po_bound} always provides
a tight bound if the used inequality holds with equality.
Still, for some games, the conditions of
Proposition~\ref{prop:po_bound_agentwise_cond} may be
easier to prove.

We now give an example of using Observation~\ref{obs:po_bound_limit}.
\setcounter{ex_match_strat}{\value{example}}
\begin{example}\label{ex:match_strat}
For any $n \geq 2$, 
consider the game with players $N = \set{1, \ldots, n}$,
the common strategy set $S_i = \set{1, \ldots, n}$,
and the utility functions
$u_i \defas \alpha_i^{\abs{\set{j \in N: s_i = s_j}}}$,
i.e.~a constant $\alpha_i$ of the player to the power of the number of
the players employing the same strategy.
Consider the solution set of all the Nash equilibria, \NE. \NE{} 
consists of all the strategy profiles where all the players employ the
same strategy. 
Since the equilibrium profiles are  exactly the profiles of maximum
social welfare, $\sum_{i = 1}^n {\alpha_i^n}$, we conclude that 
$\poa = \pos = 1$.

Let us now look at the dependency on the transition degree at transition
degree $1$. The smallest social welfare at any \NE{} is still optimal,
while the smallest social welfare at a $2$-transition set is the smallest
social welfare when agents can take any of some two strategies, and it is
$ \min_{M \subseteq N} \set{ \sum_{i \in M}{\alpha_i^{\abs{M}}} + \sum_{j \in N \setminus M}{\alpha_j^{n - \abs{M}}} }$.
Therefore, Observation~\ref{obs:po_bound_limit} implies that
$2-\pota = \poa / \frac {\sum_{i = 1}^n {\alpha_i^n}}{\min_{M \subseteq N} \set{ \sum_{i \in M}{\alpha_i^{\abs{M}}} + \sum_{j \in N \setminus M}{\alpha_j^{n - \abs{M}}} } }
= \frac {\min_{M \subseteq N} \set{ \sum_{i \in M}{\alpha_i^{\abs{M}}} + \sum_{j \in N \setminus M}{\alpha_j^{n - \abs{M}}} } }{\sum_{i = 1}^n {\alpha_i^n}}$
and $2-\pots = 1 \cdot \pos = 1$.
\end{example}

We next present the proof of Proposition~\ref{prop:po_bound_limit_agentwise_cond}.

\begin{proof}
We prove for the price of transition anarchy, and 
the proof of the price of transition stability is analogous.
First, take a least socially efficient $m$-transition, and show,
by inductive application of the $\alpha_i$-lower dependency
on a player at most $m$ times, that $u_j(s) \geq u_j(s') / (\prod_{i = 1}^{m - 1}{\alpha_i})$,
where $s'$ is some solution.
Next, being $\beta$-upper varied over $D$ implies that
$u_j(s') \geq u_j(t) / \beta$, where $t$ is a solution with the smallest
possible social welfare; perhaps, $s' = t$. Therefore,
$u_j(s) \geq u_j(t) / ((\prod_{i = 1}^{m - 1}{\alpha_i}) \beta)$, implying
that $\pota$
\begin{eqnarray*}
= \frac{\min_{s \in T(D)}{\sum_{j \in N}{u_j(s)}}}{\max_{s \in S}{\sw(s)}}
\geq \frac{{\sum_{j \in N}{u_j(t) / ((\prod_{i = 1}^{m - 1}{\alpha_i}) \beta)}}}{\max_{s \in S}{\sw(s)}}\\
= \frac{\min_{t \in D}{\sw(t)} / ((\prod_{i = 1}^{m - 1}{\alpha_i}) \beta)}{\max_{s \in S}{\sw(s)}}
= \poa / ((\prod_{i = 1}^{m - 1}{\alpha_i}) \beta).
\end{eqnarray*}
\end{proof}

We now examplify the usage of Proposition~\ref{prop:po_bound_limit_agentwise_cond}.
\setcounter{tmp}{\value{example}}
\setcounter{example}{\value{ex_match_strat}}
\begin{example}[Cont.]
For every player $j \in N$, her utility over \NE{} at
transition degree~$1$ is $\alpha_j^{n - 1}$-lower-dependent
on the transition degree, since it can drop from $\alpha_j^n$
at an \NE{} to $\alpha_j$ at $T(\NE, 2)$, if all the other players
employ another strategy. The utility is also $1$-upper-dependent
on the transition degree.
Additionally, the utility of player~$i$
is $1$-lower and -upper varied over the \NE, because in any \NE,
all the agents receive their maximum possible utility.
These two facts, used in
Proposition~\ref{prop:po_bound_limit_agentwise_cond},
employ that
$2-\pota \geq \poa / (\max\set{\alpha_j^n} \cdot 1) = \min\set{\alpha_j^{-n}}$
and $2-\pots \leq 1 \cdot 1 \cdot \pos = 1$.
\end{example}
\setcounter{example}{\value{tmp}}

We now prove Theorem~\ref{the:poa_up_bound_tight}.

\begin{proof}
The cost of any equilibrium flow is at least
the cost of letting exactly $1 / \abs{\calP_i}$ of each commodity flow
through the cheapest path, such a path costing 
$\min\set{\abs{P} : P \in \calP_i} \inf_{c \in C}{c(r_i / \abs{\calP_i})}$.
We do not have to divide $r_i$ into $\abs{\calP_i}$ equal parts, but if
a part of commodity $i$ goes through a cheaper path, than another part
of this commodity would also like to go through this part.
In the costliest transition, this commodity
all flows through the costliest path while letting it also intersect
with all the other commodities. These two ways of directing a commodity
can be in a ratio of up to $S_i(\calC)$. Therefore, the ratio of the total
flows is bounded by the maximum of $S_i(\calC)$ for all the commodities
$i = 1, \ldots, k$.

We derive \eqnsref{eq:stretch_lin} by substitution, and
to derive \eqnsref{eq:stretch_lin_no_intersect}
if the paths in different commodities never
intersect, recall that the term
$\sum_{j \in \set{1, \ldots, k} \setminus \set{i}}{r_j}$
stands for such intersections.

We show tightness by concentrating on the network 
in \figref{fig:n_par_lin}, where $a_1 > a_2 > \ldots > a_n$. Imagine such a
network for each commodity, first being separated from the networks
of other commodities.
For any $\epsilon$, we can set $a_i$s close enough to one another such that the
equilibrium flow will have values within $\epsilon / 2$-factor from one another on each edge.
Next, let $m$ such networks intersect at their topmost edges.
Then, in equilibrium, some flow of each commodity will move from the
intersection edges to the other edges, but if we pick $n$ large enough
relatively to
$m$, we can still have these flow values be within $\epsilon$ factor from $r_i / n$,
$r_i$ being the size of commodity~$i$. Therefore, the cost of the
equilibrium flow contributed by commodity~$i$ can be arbitrarily close to
${ \min\set{\abs{P} : P \in \calP_i} a_{\min}{r_i}^2 / n}$.

We now consider the transition, where all the commodities go through
their respective topmost edges. Its cost is
$\max\set{\abs{P} : P \in \calP_i} {(a_{\max} (r_i + \sum_{j \in \set{1, \ldots, k} \setminus \set{i}}{r_j}))} r_i$.
Therefore, the ratio of the two costs can be brought arbitrarily close
to $S_i(\calC)$. If all the commodities are equal, then the total ratio
of the cost of the defined transition to the cost of an equilibrium flow
is the maximum of these equal $S_i(\calC)$s, proving the tightness.
\begin{figure}[htb]
\center
\includegraphics[trim = 0mm 0mm 0mm 0mm, clip=true,width=0.31\textwidth]{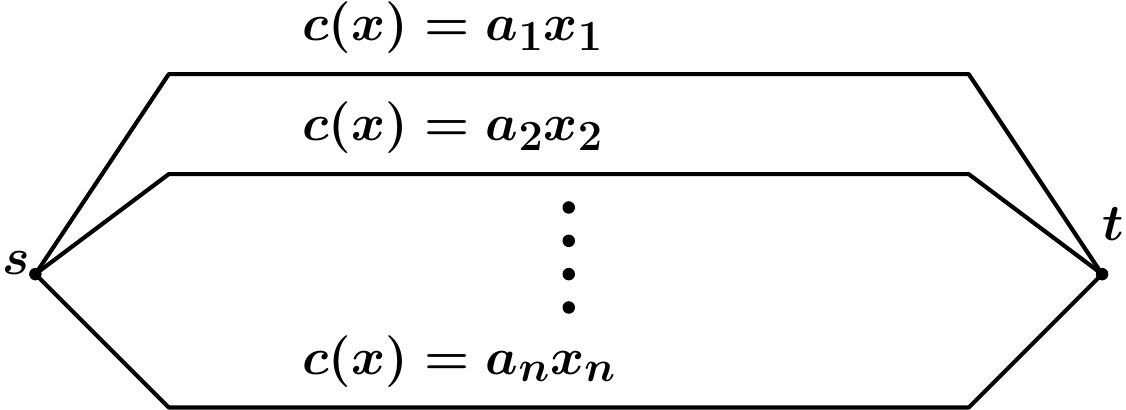}
\caption{A sub-network having $n$ parallel edges with linear costs.
}%
\label{fig:n_par_lin}
\end{figure}
\end{proof}

We now provide a proof of Proposition~\ref{prop:par_path_pots}.

\begin{proof}
Assume first that for each commodity $i$, $c_P(0) = c_{P'}(0)$ for every
$P, P' \in \calP_i$. Then, for any equilibrium flow $f$, the $f_P$ is
(strictly) positive for all paths in $\calP_i$, for the following reasons.
Consider any path $P \in \calP_i$.
The cost of any used path is positive, say $\epsilon > 0$. Since
$c_P$ is continuous, as the sum of the continuous cost functions of the
constituent edges, there exists small enough $x$, such that
$c_P(x) < \epsilon$, and so $f_P$ has to be positive.
Therefore, every feasible flow constitutes a transition, implying that
$\pots = 1$.

\end{proof}

We now prove Observation~\ref{obs:coord_game:low_sw_bound}.

\begin{proof}
If a node has strictly less than $\floor{\frac{\deg(i) - 1}{2}}$
neighbours of its colour, then it would like to deviate to the
other colour, regardless any neighbour that might change his
colour. Thus, it cannot be a stable transition.

As for \NE, having fewer than
$\ceil{\frac{\deg(i)}{2}}$ neighbours of one's colour would render
her colour suboptimal.
\end{proof}

We next prove the efficiency Theorem~\ref{The:eff_low_bound}.

\begin{proof}
The observation says that in a stable transition, the minimum
utility of player~$i$ is $\floor{\frac{\deg(i) - 1}{2}}$. In total,
the social welfare is at least
\begin{eqnarray*}
\frac{\sum_{i \in N}{\deg(i)}}{2} - \frac{\sum_{i \in N}{2}}{2}
= \frac{2 \abs{E}}{2} - \frac{2 \abs{N}}{2} = \abs{E} - \abs{N}.
\end{eqnarray*}
The maximum possible social welfare is
$
\sum_{i \in N}{\deg(i)} = 2 \abs{E},
$
implying that
$
\posta \geq \frac{\abs{E} - \abs{N}}{2 \abs{E}} = \frac{1}{2} - \frac{\abs{N}}{2 \abs{E}}.
$

The example in \figref{fig:tight_posta} demonstrates the tightness.
Indeed, the designated stable transition results in the social welfare
of~$1 + 1 + 0 + 0 + 0 = 2$, matching the bound
$\abs{E} - \abs{N} = 7 -  5 = 2$.

\begin{figure}[tb]
\center
\includegraphics[trim = 0mm 0mm 0mm 0mm, clip=true,width=0.30\textwidth]{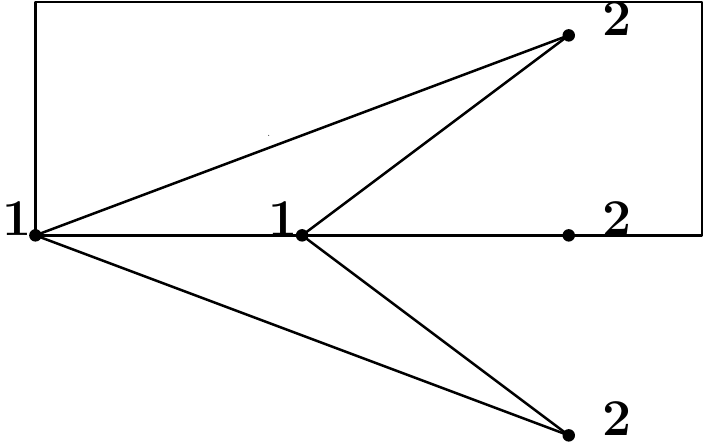}
\caption{A tight example for the price of stable transition anarchy.
}%
\label{fig:tight_posta}
\end{figure}

As the observation states, the minimum utility of a player in an \NE{}
is $\ceil{\frac{\deg(i)}{2}}$, implying the social welfare lower bound of
$
\sum_{i \in N}{\deg(i) / 2} = \abs{E}.
$
Since the maximum possible social welfare is $2 \abs{E}$, we derive
that $\poa \geq \frac{\abs{E}}{2 \abs{E}} = 1/2$.
Any cycle of an even length at least $4$ demonstrates tightness,
since the colouring $1, 1, 2, 2, \ldots, 1, 1, 2, 2$ is an \NE,
and its social welfare is $\abs{E}$, which is half the maximum.
\end{proof}

We finally prove Theorem~\ref{The:st_exist}.

\begin{proof}
For part~\ref{enum:st_exist:cycle}, consider a cycle of length $n \geq 4$.
Colour the nodes in $1$ and $2$, alternatingly; only for an odd $n$,
there will be one adjacent pair of nodes with the same colour.
This is a stable transition that is not an equilibrium. Moreover,
for and even $n$, the social welfare is $0$, exactly $\abs{E} - \abs{N}$,
attaining the lower bound from Theorem~\ref{The:eff_low_bound}.

As for part~\ref{enum:st_exist:clique}, let node~$v$ have colour~$1$.
Now, in order to have a stable transition that is not an equilibrium,
we require $v$ to have strictly
more neighbours of colour $1$, so that at least one of them is also not
best responding and its deviation will make $1$ at least as numerous
as $2$ among $v$'s neighbours.
Choosing $v$ does not restrict generality, since all the nodes are
symmetric.

If $n$ is \emph{odd}, we require $v$ to have $\frac{n - 1}{2} - 1$
neighbours coloured with $1$ and $\frac{n - 1}{2} + 1$ neighbours
coloured with $2$. Then, each node with colour $2$ will have an equal
number, namely $\frac{n - 1}{2}$, of $1$ and $2$ coloured neighbours,
and thus, every node with colour $2$ best responds, violating the
requirement for a stable transition that is not an equilibrium.

On the other hand, if $n$ is \emph{even}, we require $v$ to have
$\frac{n}{2} - 1$ neighbours coloured with $1$ and $\frac{n}{2}$
neighbours with colour $2$. Then, any node with colour $2$ has $n / 2$
neighbours of colour $1$ and $\frac{n}{2} - 1$ neighbours of colour $2$,
rendering $2$'s choice suboptimal. This implies that the
obtained profile is a stable transition though not an \NE.

Finally, we prove part~\ref{enum:st_exist:forest}.
The following definition and lemma contain the crux of the proof.
\begin{defin}
A \defined{tempered stable transition} of a coordination game
is a stable transition where some nodes have new nodes as neighbours.
Those new nodes that influence the utility of the existing nodes, but
cannot change their own colour and their utility is irrelevant.
Intuitively, they only influence the existing nodes, which allows
induction.
\end{defin}
\begin{lemma}
Any strict subtree of height at least $1$ of a coloured tree can be
re-coloured from $\set{1, 2}$ to a tempered stable transition that is not
an equilibrium.
\end{lemma}
\begin{proof}
We prove the lemma by induction on the height of the subtree.
Denote the given colour of the father~$p$ of the root~$r$ of the given
subtree with $c(p)$.

The induction basis is for height $1$. Then, colour $r$ and 
$\floor{\frac{\deg(r) - 1}{2}}$ of its children in the other
colour than $c(p)$ and colour the other children of $r$ in $c(p)$.
This is a tempered stable transition that is not
an equilibrium.

At the induction step on height $h \geq 2$, we assume the lemma for all
the subtrees of heights smaller than $h$ and prove the lemma for
a subtree of height $h$. 
Using the lemma, colour the root $r$ and all of its children
in $c(p)$.
The total colouring constitutes a tempered stable transition
that is not an equilibrium.
\end{proof}

To prove the theorem, it is enough to prove it for any tree $T$
that is not an isolated vertex.
Denote its root by $r = r(T)$ and colour $r$ in $1$.
If $T$ is of height $1$, then colour $r$ and
any $\floor{\frac{\deg(r) - 1}{2}}$ of the leaves in $1$, and let the
other leaves have colour $2$. Then, $r$ and the leaves of colour $2$
do not best respond, but if either $r$ or one of those leaves
deviates, then the other player starts best responding. Thus, this
constitutes a stable transition that is not an equilibrium.

Next, assume that $T$ is higher than $1$. First, colour $r$ and all 
the child subtrees of $r$ that are leaves in $1$. Then, colour
all the other child subtrees according to the lemma, colouring their
roots with $1$. Then, $r$ best responds, and using the lemma, we gather
the total profile is a stable transition that is not an equilibrium.
\end{proof}

%% file: decomp_zero_sum_pot.tex
\section{Decomposing into Zero-Sum and Potential Games}\label{Sec:decomp_zero_sum_pot}

Having observed that constant-sum games are insensitive to
lack of coordination and having tightly bounded the sensitivity to lack
of coordination of congestion (potential) games, we would like to
bound the sensitivity to lack of coordination of a general game that
can be decomposed to a zero-sum and a potential game.
To this end, we prove the following bound, which requires the
definition of an $\epsilon$-transition.
\begin{defin}
An \defined{$\epsilon$-$m$-transition} is an $m$-transition of the solution set
of $\epsilon$-Nash equilibria.
\end{defin}

\begin{proposition}\label{prop:game_pot_poas_potas_rat}
Let $G = (N, S, (u_i)_{i \in N})$ be a game and let 
$P = (N, S, (v_i)_{i \in N})$ be a potential game such that
$(N, S, (u_i - v_i)_{i \in N})$ is a zero-sum game%
~\cite{CandoganMenacheOzdaglarParrilo2011}.
Assume that $\forall i \in N, \abs{u_i - v_i} \leq \epsilon$ for
a positive $\epsilon$ .

If for any $2\epsilon-\NE~s$ of $P$ there exists a Nash equilibrium
$s'$ of $P$ such that $\abs{\sw(s) / \sw(s')} \geq (\leq) \alpha$,
for a fixed positive $\alpha$, then, 
\begin{eqnarray}
\abs{\frac{\poa_G}{\poa_P}} \geq \alpha, \hspace{1cm}
\paren{\abs{\frac{\pos_G}{\pos_P}} \leq \alpha}.
\label{eq:poas_ratio}
\end{eqnarray}

If for any $2\epsilon$-$m$-transition~$t$ of $P$ there exists an $m$-transition
$t'$ of $P$ such that $\abs{\sw(t) / \sw(t')} \geq (\leq) \alpha$,
for a fixed positive $\alpha$, then, 
\begin{eqnarray}
\abs{\frac{m-\pota_G}{m-\pota_P}} \geq \alpha, \hspace{1cm}
\paren{\abs{\frac{m-\pots_G}{m-\pots_P}} \leq \alpha}.
\label{eq:potas_ratio}
\end{eqnarray}
\end{proposition}
\begin{proof}
We prove \eqnsref{eq:poas_ratio} for the price of anarchy; the price of
stability result is proven similarly. Fix any Nash equilibrium $s$ of $G$.
It also
constitutes an $2\epsilon$-equilibrium of the potential game $P$, because
$\abs{u_i - v_i} \leq \epsilon$, for all agents $i$. Then, our assumption
says that there exists a Nash equilibrium
$s'$ of $P$ such that $\abs{\sw(s) / \sw(s')} \geq (\leq) \alpha$.
Therefore,
\begin{eqnarray*}
\frac{\sw_G(s)}{\max_{t \in S}{\sw_G(T)}}
\stackrel{\text{the zero-sum game does not change the social welfare}}{=} \frac{\sw_P(s)}{\max_{t \in S}{\sw_P(T)}}\\
\geq \frac{\alpha \sw_P(s')}{\max_{t \in S}{\sw_P(T)}} \geq \alpha \poa_P.
\end{eqnarray*}
Since $s$ is any \NE{} of $G$, we conclude that
$\poa_G \geq \alpha \poa_P$.

As for the price of $m$-transition anarchy and stability results, we notice
that an $m$-transition of $G$ is also a $2 \epsilon$-$m$-transition of $P$, and
from here the proof is analogous to the proof above.
\end{proof}

Having proven Theorem~\ref{Thm:cong_subadd_c_m_trans} and
Proposition~\ref{prop:game_pot_poas_potas_rat}, we conclude the following.
\begin{corollary}
Let $G = (N, S, (u_i)_{i \in N})$ be a game and let 
$P = (N, S, (v_i)_{i \in N})$ be a (utility maximization) congestion game
with superadditive utility functions such that
$(N, S, (u_i - v_i)_{i \in N})$ is a zero-sum game%
~\cite{CandoganMenacheOzdaglarParrilo2011}.
Assume that $\forall i \in N, \abs{u_i - v_i} \leq \epsilon$ for
a positive $\epsilon$ .

If for any $2\epsilon$-$m$-transition~$t$ of $P$ there exists an $m$-transition
$t'$ of $P$ such that $\abs{\sw(t) / \sw(t')} \geq \alpha$,
for a fixed positive $\alpha$, then, 
$$
m-\pota_G \geq \frac{\alpha}{m} \poa_P.
$$
\end{corollary}
\begin{proof}
Theorem~\ref{Thm:cong_subadd_c_m_trans} implies $m-\pota_P \geq \poa_P / m$,
because we now talk about utilities rather than costs.
Then, \eqnsref{eq:potas_ratio} implies that
$m-\pota_G \geq \alpha m-\pota_P$, and therefore,
$m-\pota_G \geq \alpha m-\pota_P \geq \alpha \poa_P / m$.
\end{proof}